\documentclass[12pt]{article}

\usepackage{amsmath}
\usepackage{graphicx,psfrag,epsf}
\usepackage{enumerate}
\usepackage{natbib}
\usepackage{url} % not crucial - just used below for the URL 

\newcommand{\blind}{1}

\usepackage{psfrag,epsf}
\usepackage{enumerate}
\usepackage{colortbl}

\usepackage{tikz}
\usetikzlibrary{angles,quotes}
\usetikzlibrary{shapes.geometric, arrows}

\tikzstyle{startstop} = [rectangle, rounded corners, minimum width=3cm, minimum height=1cm,text centered, draw=black, fill=yellow!60]
\tikzstyle{io} = [rectangle, rounded corners, minimum width=3cm, minimum height=1cm, text centered, draw=black, fill=yellow!60]
\tikzstyle{process} = [rectangle, rounded corners, minimum width=3cm, minimum height=1cm, text centered, text width=4cm, draw=black, fill=yellow!60]
\tikzstyle{method1} = [rectangle, rounded corners, minimum width=3cm, minimum height=1cm, text centered, text width=4cm, draw=black, fill=red!60]
\tikzstyle{method2} = [rectangle, rounded corners, minimum width=3cm, minimum height=1cm, text centered, text width=4cm, draw=black, fill=black!20!green]
\tikzstyle{method3} = [rectangle, rounded corners, minimum width=3cm, minimum height=1cm, text centered, text width=4cm, draw=black, fill=blue!50]
\tikzstyle{arrow} = [thick,->,>=stealth]

\def\boxit#1{\vbox{\hrule\hbox{\vrule\kern6pt
          \vbox{\kern6pt#1\kern6pt}\kern6pt\vrule}\hrule}}

\def\wh{\widehat}

\def\log{\hbox{log}}

\def\Normal{\hbox{Normal}}

\def\bse{\begin{eqnarray*}}
\def\ese{\end{eqnarray*}}
\def\be{\begin{eqnarray}}
\def\ee{\end{eqnarray}}
\def\bq{\begin{equation}}
\def\eq{\end{equation}}
\def\bse{\begin{eqnarray*}}
\def\ese{\end{eqnarray*}}

\def\wh{\widehat}
\def\trans{^{\rm T}}

\def\b1e{{\mathbf e}}

\def\bS{{\mathbf S}}
\def\bzero{{\mathbf 0}}

\newcommand{\bgamma}{\mbox{\boldmath $\gamma$}}
\newcommand{\bzeta}{\mbox{\boldmath $\zeta$}}
\newcommand{\bSigma}{\mbox{\boldmath $\Sigma$}}

\def\bbeta{{\boldsymbol{\beta}}}
\def\bmu{{\boldsymbol{\mu}}}

\def\btheta{{\boldsymbol{\theta}}}
\def\bgamma{{\boldsymbol{\gamma}}}

\def\bzero{{\boldsymbol{0}}}
\def\bone{{\boldsymbol{1}}}

\def\bb{{\boldsymbol b}}

\def\bg{{\boldsymbol g}}
\def\bh{{\boldsymbol h}}

\def\bk{{\boldsymbol k}}

\def\bq{{\boldsymbol 	q}}
\def\br{{\boldsymbol r}}
\def\bs{{\boldsymbol s}}

\def\by{{\boldsymbol y}}

\def\bD{{\boldsymbol D}}

\def\bI{{\boldsymbol I}}

\def\bM{{\boldsymbol M}}
\def\bN{{\boldsymbol N}}

\def\bP{{\boldsymbol P}}

\def\bR{{\boldsymbol R}}
\def\bS{{\boldsymbol S}}
\def\bT{{\boldsymbol T}}

\def\bV{{\boldsymbol V}}

\def\bY{{\boldsymbol Y}}
\def\bZ{{\boldsymbol Z}}

\def\boxit#1{\vbox{\hrule\hbox{\vrule\kern6pt\vbox{\kern6pt#1\kern6pt}\kern6pt\vrule}\hrule}}

\def\tilD{\bD}

\def\ACEest{\wh{\btheta}_{\rm ACE}}
\def\E{\rm E}
\def\Var{\rm Var}

\newtheorem{Th}{\underline{\bf Theorem}}

\newtheorem{proposition}{Proposition}
\newtheorem{Lem}{\underline{\bf Lemma}}

\begin{document}

\def\spacingset#1{\renewcommand{\baselinestretch}%
{#1}\small\normalsize} \spacingset{1}

%%%%%%%%%%%%%%%%%%%%%%%%%%%%%%%%%%%%%%%%%%%%%%%%%%%%%%%%%%%%%%%%%%%%%%%%%%%%%%

\if1\blind
{
  \title{\bf Mission Imputable: Correcting for Berkson Error When Imputing a Censored Covariate}
  \author{Kyle F. Grosser\thanks{
    The authors gratefully acknowledge \textit{the National Institute of Environmental Health Sciences grants T32ES007018 and P30ES010126 and the National Institute of Neurological Disorders and Stroke grant K01NS099343.}}\hspace{.2cm}\\
    Department of Biostatistics, University of North Carolina \\ at Chapel Hill Gillings School of Global Public Health\\
    Sarah C. Lotspeich \\
    Department of Statistical Sciences, Wake Forest University \\
    Tanya P. Garcia \\
    Department of Biostatistics, University of North Carolina \\ at Chapel Hill Gillings School of Global Public Health}
  \maketitle
} \fi

\if0\blind
{
  \bigskip
  \bigskip
  \bigskip
  \begin{center}
    {\LARGE\bf Mission Imputable: Correcting for Berkson Error When Imputing a Censored Covariate}
\end{center}
  \medskip
} \fi

\bigskip
\begin{abstract}
To select outcomes for clinical trials testing experimental therapies for Huntington disease, a fatal neurodegenerative disorder, analysts model how potential outcomes change over time. Yet, subjects with Huntington disease are often observed at different levels of disease progression. To account for these differences, analysts include time to clinical diagnosis as a covariate when modeling potential outcomes, but this covariate is often censored. One popular solution is imputation, whereby we impute censored values using predictions from a model of the censored covariate given other data, then analyze the imputed dataset. However, when this imputation model is misspecified, our outcome model estimates can be biased. To address this problem, we developed a novel method, dubbed ``ACE imputation.'' First, we model imputed values as error-prone versions of the true covariate values. Then, we correct for these errors using semiparametric theory. Specifically, we derive an outcome model estimator that is consistent, even when the censored covariate is imputed using a misspecified imputation model. Simulation results show that ACE imputation remains empirically unbiased even if the imputation model is misspecified, unlike multiple imputation which yields $>100\%$ bias. Applying our method to a Huntington disease study pinpoints outcomes for clinical trials aimed at slowing disease progression.
\end{abstract}

\noindent%
{\it Keywords:} Censored data, Huntington disease, imputation correction, measurement error, semiparametric theory
\vfill

\newpage
\spacingset{1.45} % DON'T change the spacing!

\section{Introduction}
\label{sec:intro}
\subsection{Statistical hurdles for clinical trials of Huntington disease}
Clinical trials are now underway to test experimental therapies aimed at slowing the progression of Huntington disease, a genetically inherited disease that leads to progressive cognitive and motor impairment. No effective therapies for the disease have been developed yet; one major difficulty is finding an outcome to help assess if an experimental therapy has an effect \citep{langbehn2020clinical}. Clinical trialists want to avoid selecting an outcome that changes too slowly over time, because then it would be difficult to determine if an experimental therapy actually slowed progression. This is because, for an outcome that changes slowly, the progression would not have changed much over the trial period regardless of the therapy being tested.

\begin{figure}[!t]
    \centering
    \includegraphics[width=0.8\textwidth]{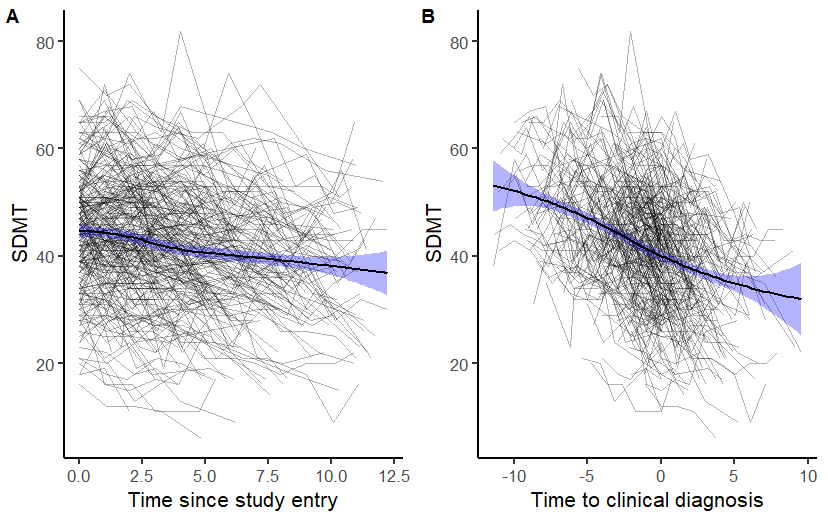}
    \caption{Symbol Digit Modality Test (SDMT) scores from uncensored (i.e., clinically diagnosed) subjects plotted against (\textbf{A}) time since study entry and (\textbf{B}) time to clinical diagnosis. Thin lines indicate change over time for individual subjects, and the thick line indicates a LOESS curve summarizing the overall trajectories.}
    \label{fig:compare_s_and_s_T_uncens}
    % take out the T* plots, remove grid, are the numbers readable?, move the T* plots to later section, think about coloring, remove color (needs to be black and white ready)
\end{figure}

Therefore, an ideal outcome would be one %Finding the best outcome (i.e., an outcome 
in which change can be easily detected over the course of the trial, but %to use in a clinical trial
determining such an outcome is statistically challenging. Data measuring cognitive and motor impairment over time are typically collected from individuals who are at different levels of disease progression, some more advanced than others. These differences can obscure our search to identify the best outcome to use in a clinical trial. For example, plots of the Symbol Digit Modalities Test (SDMT) scores (a measure of cognitive impairment) show little to no change over the course of the study if we do not adjust for individuals being at different levels of disease progression (Figure~\ref{fig:compare_s_and_s_T_uncens}A). To capture disease progression, we use time \textit{to} clinical diagnosis, which refers to the difference between the time at which a subject is observed and their time \textit{of} clinical diagnosis (i.e., the day on which a clinician determines that a subject's motor impairment is unequivocally attributable to Huntington disease) \citep{kieburtz1996controlled}. % (Section~\ref{sec:notation} demonstrates this distinction using math). 
Now, %if we do make that adjustment, such as 
by plotting SDMT scores as a function of time to clinical diagnosis instead of simply time in the study (i.e., adjusting for disease progression), we see that cognitive impairment worsens rapidly in the time period immediately before and after diagnosis (Figure~\ref{fig:compare_s_and_s_T_uncens}B). These findings based on the adjusted trajectories of cognitive impairment %This rapid worsening in cognitive impairment %--- which shines through when we include time to clinical diagnosis --- 
agree with previous clinical findings, whereas those based on the unadjusted trajectories %barely seeing any changes does 
do not \citep{paulsen2008detection}.

Modeling cognitive impairment as a function of time to clinical diagnosis is a common way to adjust for individuals being at different levels of disease progression \citep{long2014tracking}. With data from a fully diagnosed cohort, making this adjustment to our analyses would be straightforward. However, because Huntington disease progresses slowly over time, time of clinical diagnosis is often unobserved for at least some subjects in a given cohort. Yet, Huntington disease is fully penetrant, i.e., anyone who inherits the genetic mutation should eventually meet the criteria for a diagnosis. We therefore know that, for undiagnosed subjects, the time of clinical diagnosis must lie beyond when they were last observed. This phenomenon is known as right censoring. Hence, to select outcomes for clinical trials of Huntington disease, we must accurately model the change in potential outcomes given a right-censored covariate: time of clinical diagnosis.

\subsection{Statistical modeling with a censored covariate}

One might be tempted to replace censored subjects' times of clinical diagnosis with their times of last observation instead (e.g., last visit), and then model the potential outcomes using these replacements. However, %because of the right censoring that is inherent to our problem, 
these right-censored replacements will be necessarily less than the true clinical diagnosis times. As a result, such ``naive'' analyses tend to produce biased model estimates \citep{austin2003type}.
Alternatively, we could remove the censored subjects from our data, and then model the potential outcomes using only the uncensored data. % \cite{maroof2011modeling} adopted this approach in their analysis of motor and cognitive changes in subjects at risk for Huntington disease. 
However, it is well established that such ``complete case'' analyses yield parameter estimates that are less precise (i.e., less efficient) than those that would be obtained if the full dataset were available \citep{ALA}. %This is because by using only complete cases, we are discarding data and therefore reducing our effective sample size. 
Complete case analyses can also lead to biased parameter estimates and inflated type I error rates for hypothesis tests \citep{austin2003type}. % When \cite{maroof2011modeling} used complete case analysis in their analysis of Huntington disease data, this approach left them with a sample size of only 19.

Hence, we do not want to ``throw away'' our uncensored data as in complete case analyses, but we cannot use the data ``as is'' as in naive analyses. Imputation --- whereby we impute censored covariate values and then analyze the imputed dataset --- provides a more promising solution. 
% Several imputation methods have been developed to address censored covariates. 
% To perform imputation, we replace (i.e., ``impute'') censored covariate values. 
%These replacements can be generated in many ways, including drawing predictions from a model of the censored covariate given other, fully observed data (i.e., an ``imputation model''). \cite{Bernhardt2014} and \cite{Wei2018} proposed a fully parametric model of the censored covariate given fully observed data. \cite{Atem2019} and \cite{correctingCMI} offered a more flexible, semiparametric approach, modeling the hazard function of the censored covariate with a Cox proportional hazards model, then using this hazard model as the imputation model. \cite{Wang2012} and \cite{yu2021quantile} introduced another semiparametric approach, where the imputation model is a quantile regression model of the censored covariate. 
These imputed values can be generated from a model of the censored covariate given other, fully observed data (an ``imputation model'') in many ways, including random draws \citep{Bernhardt2014, Wei2018}, conditional means \citep{Atem2019}, and conditional quantiles \citep{Wang2012, yu2021quantile}. 

Although imputation for censored covariates is still a growing area of research, we can learn much from the massive body of literature concerning imputation for ``traditional'' missing data. Importantly, it is well understood that imputation for traditional missing data will yield consistent estimates of our outcome model parameters given two key assumptions: i) the data are missing at random and ii) the imputation model is correctly specified \citep{little2019statistical}. In particular, the missing at random assumption requires that the probability of the variables ``being missing'' is conditionally independent of the missing variables themselves, given the other fully observed variables.
% It is well understood that if two key assumptions hold, then imputation will yield consistent parameter estimates when applied to traditional missing data. These assumptions are 1) that data are missing at random and 2) that the imputation model is correctly specified. The missing at random assumption requires that the probability of being missing is conditionally independent of the (possibly) missing variables, given the fully observed variables (i.e., variables which are observed for all subjects) in our data. 
With censored data, this assumption is immediately violated since the probability that a variable $X$ is right censored by a variable $C$ (i.e., $P(C<X)$) depends directly on $X$. For example, with right-censored data, a possibly censored variable $X$ is more likely to be censored by a variable $C$ if $X=2$ than if $X=4$, since $P(C<4)\leq P(C<2)$ (with strict inequality in general). Fortunately, \cite{Bernhardt2014} and \cite{Wang2012} show (both in theory and simulation) that consistent parameter estimates can still be obtained when we use imputation to address censored covariates. These findings demonstrate that the missing at random assumption is not a necessary assumption when imputing censored covariates. Yet, all of the cited imputation approaches hinge on a model of the censored covariate; when this imputation model is misspecified, it could introduce bias into our outcome model estimates \citep{yucel2010impact, black2011missing}. 
% To our knowledge, there does not yet exist a method to account for imputation model misspecification.

\subsection{Overview}

In this paper, we present a novel method to consistently estimate a linear mixed effects model for a continuous outcome given a censored covariate. We build on existing imputation methods for censored covariates, which result in biased outcome model estimates when the imputation model is misspecified, by accounting for such misspecification in order to reduce that bias.
% to consistently model a longitudinal outcome given a censored covariate imputed with its conditional mean, even when the imputation model is misspecified. % First, we implement the semiparametric imputation method described by \cite{Atem2019} and \cite{correctingCMI} to impute the censored covariate.
To account for possible imputation model misspecification, we model the difference between the true (but censored) covariate values and their imputed replacements using a Berkson error model \citep{Carrolletal2006}. Then, to avoid possibly misspecifying the distribution of this ``imputation error'' (since that could lead to bias and inefficiency), we adapt a flexible, semiparametric method developed by \cite{garcia2016optimal} that allows the imputation error to follow any distribution. 
We introduce our proposed method, ``active correction for error in imputation'' (dubbed ``ACE imputation''), in Section~\ref{sec:methods}.  
In Section~\ref{sec:ee-properties}, we show that ACE imputation produces an estimator that is identifiable, consistent, and asymptotically normal, even under imputation model misspecification.
In  Section~\ref{sec:simulation}, simulations show that ACE imputation outperforms competing solutions for censored covariates when the imputation model is correct and remains unbiased even when the underlying imputation model has been misspecified. 
In Section~\ref{sec:real_data_analysis}, we show that applying ACE imputation to Huntington disease data %from PREDICT-HD 
 %, a longitudinal study of Huntington disease, 
helps pinpoint outcomes %which change quickly over time so that these outcomes can be used 
that could be used to test experimental therapies for Huntington disease. Section~\ref{sec:discussion} concludes this paper with a discussion of potential limitations and future work.
% In this paper, we present a novel method to estimate a longitudinal outcome model in the presence of a censored covariate. We first implement the semiparametric imputation method described by \cite{Atem2019} and \cite{correctingCMI} to impute values for the censored covariate; then, we incorporate these imputed values into a novel estimating equation that accurately estimates our model of interest \textit{even when the underlying covariate model is incorrect.} 
% Brag about our method without introducing a bunch of new terminology

% The rest of this paper is organized as follows. We describe our novel method in Section~\ref{sec:methods} then present its theoretical properties in Section~\ref{sec:ee-properties}. In Section~\ref{sec:simulation}, we demonstrate our method's performance in simulation, comparing it to competing solutions for censored covariates. In Section~\ref{sec:real_data_analysis}, we use our method to compare possible outcomes for Huntington disease trials, based on how fast these outcomes change in untreated subjects.

\section{Methods}
\label{sec:methods}
To better pinpoint outcomes for clinical trials of Huntington disease (i.e., outcomes that change quickly enough so that effective therapies can be reliably identified) we must accurately model how potential outcomes progress over time as a function of time of clinical diagnosis, a right-censored covariate. In this section, we present our ACE imputation method, which estimates the parameters of a longitudinal model given a right-censored covariate. To achieve this goal, we first implement an existing solution to covariate censoring, in which we impute censored values from a Cox model. Then, because this method can produce bias when that imputation model is misspecified, we reduce this bias by adjusting for the errors that occur due to such misspecification. 

\subsection{Notation and longitudinal model}
\label{sec:notation}
We longitudinally model our potential outcomes using data collected for $n$ subjects (indexed by $i$, $i=1,\ldots,n$) and $m_i$ observations per subject (indexed by $j$, $j=1,\ldots,m_i$). The outcome for subject $i$ at visit $j$ is $Y_{ij}$, observed at time $s_{ij}$. We also observe time of clinical diagnosis $X_i$, $p_a$-dimensional covariates $\bZ^a_{ij}$, and $p_b$-dimensional covariates $\bZ^b_{ij}$. To account for clustering between outcomes from the same subject, we include a $p_b$-dimensional vector of unobserved, subject-specific random effects $\bb_i$. In Huntington disease, for example, $Y_{ij}$  is the outcome being considered for a clinical trial, $s_{ij}$ is the time of observation (years) since the start of the study, $X_i$ is the time of clinical diagnosis (years) relative to the start of study, $\bZ^a_{ij}$ are baseline age, sex, education, and genetic information, and $\bb_i$ is a subject-specific random intercept (hence, $\bZ^b_{ij}$ is simply a vector of ones).

To account for clustering between $Y_{i1},\ldots,Y_{im_i}$, we employ the linear mixed model:
\bse
Y_{ij} =
\alpha(s_{ij} - X_i) + \bbeta^T \bZ^a_{ij} + \bb_i^T \bZ^b_{ij} + \epsilon_{ij}, \ \ \ \bb_i\sim f_{\bb}, \ \ \ \epsilon_{ij}\sim\Normal(0,\sigma^2).\ese
Herein, $\alpha$ is the parameter associated with $(s_{ij} - X_i)$, $\bbeta$ is the $p_a$-dimensional parameter vector associated with $\bZ^a_{ij}$, and $\sigma^2$ is the variance of the random errors $\epsilon_{ij}$.

We assume that subjects are independent of one another and that outcomes $Y_{ij}$ from the $i$th subject are conditionally independent given $\bb_i\sim f_{\bb}$. Typically with mixed models, analysts assume a distribution for $\bb$ such as a multivariate normal distribution with mean $\bzero$ and unknown covariance matrix. To promote model flexibility, we instead allow $\bb$ to follow any distribution $f_{\bb}$. We do this so to avoid possibly misspecifying $f_{\bb}$, since misspecification can lead to bias and inefficiency \citep{garcia2016optimal}.

We refer to $(s_{ij} - X_i)$ as ``time to clinical diagnosis'' (whereas $X_i$ on its own is time \textit{of} clinical diagnosis). To illustrate, $(s_{ij} - X_i) = -2$ when subject $i$ is two years before they are clinically diagnosed and $(s_{ij} - X_i) = 2$ when they are two years after being clinically diagnosed. As discussed in Section~\ref{sec:intro}, we use this difference rather than just $s_{ij}$ to capture the fact that some subjects are farther along in their disease progression than others.
A key challenge with the model is that $X_i$ is often right-censored; in the Huntington disease data that we analyze, diagnosis time $X_i$ is right-censored for $\approx 78\%$ of subjects. Hence, rather than observe $X_i$, we observe $W_i=\min(X_i,C_i)$ and $\Delta_i=I(X_i\leq C_i)$, where $C_i$ is a continuous, random right-censoring time and $\Delta_i$ is the censoring indicator. We refer to $C_i$ as a ``right-censoring'' time because when time of clinical diagnosis $X_i$ is censored by $C_i$, we know that $X_i$ falls to the right of $C_i$, i.e., $X_i > C_i$.

In summary, the observed data are $(\bY_i,\bs_i,\bZ^a_i,\bZ^b_i,W_i,\Delta_i)$ for $i=1,\ldots,n$ where  $\bY_i=(\bY_{i1},\ldots,\bY_{im_i})^T,\bs_i=(s_{i1},\ldots,s_{im_i})^T,\bZ^a_i=(\bZ^a_{i1},\ldots,\bZ^a_{im_i})^T $, and $\bZ^b_i=(\bZ^b_{i1},\ldots,\bZ^b_{im_i})^T$.
We want to estimate  $\btheta=(\alpha, \bbeta\trans, \sigma^2)\trans$ in the presence of right-censoring on $X_i$.

\subsection{Imputing censored times of diagnosis}
\label{sec:cmi}
% Although one might be tempted to fit the mixed models described in Section~\eqref{sec:notation} with $X_i$ replaced by $W_i=\min(X_i,C_i)$ for all subjects, this can lead to biased parameter estimates \citep{austin2004estimating}. This is because for censored subjects ($i$ such that $\Delta_i=0$), $W_i$ will be less than $X_i$, and so such a \textit{naive analysis} can yield bias. Alternatively, we could analyze data only from subjects with observed times of clinical diagnosis $X_i$ ($i$ such that $\Delta_i=1$) and discard data from those without. This approach, known as \textit{complete case analysis}, is simple to implement but, by discarding data from censored subjects, we are reducing the sample size of our data, which leads to more variable (i.e., less efficient) parameter estimates \citep{ALA}.
% Imputation is a more compelling alternative because it allows us to ``fill in'' censored times of clinical diagnosis, yielding a newly complete dataset that we can then analyze with traditional methods and software \citep{rubin1988overview}. Furthermore, by preserving data from \textbf{all} subjects, we are increasing the precision of our parameter estimates relative to complete case analysis. Analyzing an imputed dataset will not yield the same level of efficiency as analyzing a dataset which was never censored to begin with, but it is more efficient than outright discarding censored subjects.
Imputation is a compelling solution to covariate censoring because it allows us to ``fill in'' censored times of clinical diagnosis. We can then analyze this newly complete dataset using traditional methods such as restricted maximum likelihood estimation (REML). Furthermore, by preserving data from all subjects, we increase the efficiency of our parameter estimates relative to complete case analysis. Analyzing an imputed dataset will not yield the same level of efficiency as analyzing a dataset that was never censored to begin with, but it is more efficient than discarding censored subjects outright.

% Benefits for imputation:
% 1. Increased precision relative to complete case analysis
% 2. Creates a 'completed' dataset that can then be analyzed with "off the shelf" methods

% FOR IMPUTATION OF GENERAL MISSING DATA (not censored)
% 3. Many available methods to do imputation.
% 4. When the imputation model is correctly specified, MI produces consistent and asymptotically normal results

% SPECIFIC TO IMPUTATION OF CENSORED DATA:
% 5. Less bias than a naive analysis (using W = min(T, C) in place of T)
% 6. Cox-based CMI has a straightforward formula (easy to implement with our package!).
% 7. With Cox-based CMI, we are able to impute values that are guaranteed to be > C

% Therefore, to estimate $\btheta$ in the presence of covariate censoring, we first impute censored values of $X_i$. Although there have been many imputation methods developed to handle traditional missing data \citep{murray2018multiple}, we must tailor our approach to account for the right-censoring of $X_i$ by $C_i$. This is because we know that $X_i > C_i$ due to right-censoring, and so we should only impute values which are greater than $C_i$. To do so, we use conditional mean imputation, conditioning on our a priori knowledge that $X_i > C_i$ for censored subjects. More specifically, we employ conditional mean imputation based on a Cox model \citep{Atem2019, correctingCMI}, which adjusts for censoring and does not impose strict distributional assumptions on the time of clinical diagnosis $X_i$; this is appealing because relying on incorrect distributional assumptions can yield invalid inference.

Therefore, to estimate $\btheta$ in the presence of covariate censoring, we first impute censored values of $X_i$. 
% Although there have been many imputation methods developed to handle traditional missing data \citep{murray2018multiple}, we must tailor our approach to account for the right-censoring of $X_i$ by $C_i$. This is because we know that $X_i > C_i$ due to right-censoring, and so we should only impute values which are greater than $C_i$. To do so, we use conditional mean imputation, conditioning on our \textit{a priori} knowledge that $X_i > C_i$ for censored subjects. 
More specifically, we employ conditional mean imputation based on a Cox model \citep{Atem2019, correctingCMI}, which adjusts for censoring and does not impose strict distributional assumptions on the time of clinical diagnosis $X_i$. First, we fit a Cox model with times of diagnosis, $X_i$, as the outcome given a vector of time-invariant covariates $\bV_i$. The covariates $\bV_i$ can be a subset of the variables in $\bZ^a_i$ or $\bZ^b_i$, or auxiliary variables. 
% \textbf{Therefore, in the first stage of our two-stage estimation procedure, we implement Cox-based conditional mean imputation.} To do so, we first fit a Cox model with times of clinical diagnosis, $X_i$, as the outcome given the covariate vectors $\bV_i$ for subjects $i=1,\ldots,n$. 
The Cox model assumes that the hazard function for $X_i$ given $\bV_i$ is
\bse
\lambda(x|\bV_i)=\lambda_0(x)\exp(\bgamma^T\bV_i),
\ese
where $\lambda_0(x)$ and  $\bgamma$ are the baseline hazard function and covariate effects, respectively. The elements of $\bgamma$ represent log-hazard ratios for the corresponding elements of $\bV_i$. After fitting this model, we replace times of clinical diagnosis $X_i$ with: 
\be
\label{eqn:def_X_i_star}
\widehat{X}_i&=&\Delta_iX_i+(1-\Delta_i){\rm E}(X_i|X_i>C_i,\bV_i).
\ee
For uncensored subjects, $X_i$ is unchanged (i.e., $\widehat{X}_i=X_i$ for $i$ such that $\Delta_i = 1$). For censored subjects, we replace $X_i$ with its conditional mean given $X_i>C_i$ and $\bV_i$ (i.e., $\widehat{X}_i={\rm E}(X_i|X_i>C_i,\bV_i)$ for $i$ such that $\Delta_i = 0$). 
% \cite{correctingCMI} show that this conditional mean is equivalent to:
% \bse
% {\rm E}(X_i|X_i>C_i,\bV_i)&=&C_i+\frac{\int_{C_i}^{\infty}\{S_{0}(t)\}^{\exp(\wh \bgamma^T\bV_i)}dt}{\{S_{0}(t)\}^{\exp(\wh\bgamma^T\bV_i)}}
% \ese
% where $S_0(v|\bV_i)$ is the baseline survival function for $X_i$ adjusted for covariates $\bV_i$. 
\cite{correctingCMI} show that this conditional expectation can be approximated as $\widehat{\rm E}(X_i|X_i>C_i,\bV_i) =$
\bse
& C_i + \frac{1}{2}\left[\frac{\sum_{j=1}^{n-1}{\rm I}(W_{(j)} \geq C_i)\left\{S_{0}\left(W_{(j+1)}\right)^{\exp(\bgamma^T\bV_i)}+ S_{0}\left(W_{(j)}\right)^{\exp(\bgamma^T\bV_i)}\right\}\left(W_{(j+1)}-W_{(j)}\right)}{S_{0}(C_i)^{\exp(\bgamma^T\bV_i)}}\right],
\ese
where $W_{(1)} < W_{(2)} < \dots < W_{(n)}$ are the ordered values of $W = \min(X, C)$ and $S_0(x|\bV_i)$ is the baseline survival function for $X_i$ given $\bV_i$. To compute this approximation, we estimate the log-hazard ratios $\bgamma$ using standard statistical software and the baseline cumulative hazard function $S_0(\cdot)$ using the Breslow estimate \citep{breslow1972contribution}.

\subsection{Why imputation introduces errors}
\label{sec:berkson}
For imputation to yield consistent parameter estimates, we must correctly specify this Cox model. When this model is misspecified, the imputed values $\widehat{X}_i$ can be very far from the true times of clinical diagnosis, $X_i$. As we empirically show in Section~\ref{sec:simulation}, this ``imputation error'' (i.e., the difference between $X_i$ and $\widehat{X}_i$) can lead to serious bias.
% To our surprise, we found that even when the imputation model has been \textit{properly} specified, bias can still occur. 

To reduce this bias, we model the imputation error. \cite{haber2020bias} note that imputation error is a type of Berkson error, which arises when subjects in a group are assigned the same value for a missing variable. When subjects with similar traits are all assigned the same value (e.g., an estimated group mean), error arises because true individual values deviate from this group mean by an unobserved amount. With conditional mean imputation, we replace censored $X_i$ with $\widehat{X}_i={\rm E}(X_i|X_i>C_i,\bV_i)$. So, if two subjects $i$ and $j$ have the same traits, i.e., $(C_i, \bV_i) = (C_j, \bV_j)$, we will assign equivalent imputed values to each, i.e., $\widehat{X}_i = \widehat{X}_j$. As a result, we know that the relationship between $X_i$ and $\widehat{X}_i$ follows a Berkson error model. We represent this as
\bse
X_i = \widehat{X}_i + U_i
\ese
with measurement error $U_i\sim f_U$ \citep{Carrolletal2006}. 
Given the definition of $\widehat{X}_i$ in Equation~\eqref{eqn:def_X_i_star}, there is no imputation error when a subject's time of clinical diagnosis is uncensored (i.e., $U_i=0$ for $i$ such that $\Delta_i=1$), since we need not impute for that subject.
% This led us to construct the second stage of our two-stage estimation procedure, where we account for this imputation error in order to accurately estimate $\btheta$, even when our imputation model has been misspecified.
\subsection{Correcting for errors in imputation}
\label{sec:Hilbert-space-stuff}
Since we know that error between the true (but censored) values of $X_i$ and their imputed replacements $\widehat{X}_i$ can lead to bias (Section~\ref{sec:simulation}), we adjust for this imputation error ($U_i$) to more accurately estimate $\btheta$. %As with the random effects $\bb_i$, we do not make any distributional assumptions about $U_i$, and therefore leave its probability density function $f_{U_i}(\cdot)$ completely unspecified. We do this so as to avoid the risk of possibly misspecifying these components of the model, which may lead to invalid inference. %\cite{Tsiatis2006} presents a technique for deriving an estimating equation to estimate parameters of interest (e.g., $\btheta$) in the presence of such ``nuisance'' distributions (e.g., $(f_{\bb}, f_{U})$).
To estimate $\btheta$  under the most flexible modeling assumptions, we will not make any distributional assumptions on the random effects, $\bb_i$, or imputation error, $U_i$. Instead, treating the random effects and imputation error as so-called latent variables (i.e., variables not observed), our longitudinal mixed effects model belongs in the class of generalized linear latent variable models. For such a class of models, a semiparametric method exists \citep{garcia2016optimal} for which model parameters are estimated using an intermediate quantity that plays a similar role as that of the classical sufficient and complete statistic.
This method results in parameter estimators that are consistent, efficient, and robust to misspecification. 

%To construct this estimating equation, we must choose appropriate estimating functions. Hence, we first consider the space of functions with mean $\bzero$ (so that these functions can be used as the estimating functions of an estimating equation). Ideally, our estimating functions would be impacted very little by misspecification of our nuisance distributions. To find estimating functions with this property, we first define the set of possible estimating functions which are maximally affected by possible misspecification of $(f_{\bb}, f_{U})$. We then use this set of functions ---- which we denote $\Lambda$ --- to define its orthogonal complement, $\Lambda^{\perp}$. This is the set of functions that are impacted as little as possible --- or, perhaps, completely unaffected --- by $(f_{\bb}, f_{U})$. We want our estimating equation to have this property, and so we consider estimating functions belonging to $\Lambda^{\perp}$ as candidates.

Extending this semiparametric method to our problem, we show in Appendix~\ref{sec:original_score_vectors} that the estimator of interest, $\ACEest$, is the solution to  the estimating equation:
\be
\label{eqn:total_estimating_equation}
\sum_{i=1}^n \{\bS_{\rm{eff}}^{\rm{full}}(\bY_i,\tilD_i,\widehat{\tilD}_i;\btheta)\}\equiv
\sum_{i=1}^n \left\{\Delta_i\bS_{\rm{eff}}(\bY_i,\tilD_i;\btheta) + (1 - \Delta_i)\bS_{\rm{eff}}^*(\bY_i,\widehat{\tilD}_i;\btheta)\right\}=\bzero,
\ee
where $\bS_{\rm{eff}}(\bY_i,\tilD_i;\btheta)$ is the score vector for uncensored data 
$\tilD_i=(\bs_i, X_i, \bZ^a_i, \bZ^b_i)$,  and 
$\bS_{\rm{eff}}^*(\bY_i,\widehat{\tilD}_i;\btheta)$ is the score vector for censored data $\widehat{\tilD}_i=(\bs_i, \widehat{X}_i, \bZ^a_i, \bZ^b_i)$. The derivation of these score vectors shows that initially, constructing these score vectors requires solving a computationally slow and numerically unstable problem. However, by leveraging properties of multivariate normal distributions and linear projection theory, we derive straightforward, closed-form versions of the score vectors (see Appendix~\ref{sec:simplify_score_vector}).

In Appendix~\ref{sec:calcs_uncens}, we show that $\bS_{\rm{eff}}(\bY_i,\tilD_i;\btheta)=$
\bse
\label{eqn:uncens_eff_score_vec}
\sigma^{-4}\left[
\begin{matrix}
\sigma^2\bZ^{aT}_i(\bI_{m_i} - \bP_{\bZ^b_i})\{\bY_i-\bzeta(\tilD_i;\btheta)\} \\
\sigma^2(\bs_i-X_i\bone_{m_i})^T(\bI_{m_i} - \bP_{\bZ^b_i})\{\bY_i-\bzeta(\tilD_i;\btheta)\} \\
\frac{1}{2}\{\bY_i^T\bY_i-{\rm E}(\bY_i^T\bY_i|\bT_i,\tilD_i)\}-\bzeta^T(\tilD_i;\btheta)\{\bY_i-{\rm E}(\bY_i|\bT_i,\tilD_i)\}
\end{matrix}
\right],
\ese
where
\bse
{\rm E}(\bY_i|\bT_i,\tilD_i)&=&\bP_{\bZ^b_i}\bY_i+(\bI_{m_i}-\bP_{\bZ^b_i})\bzeta(\tilD_i;\btheta)\\
{\rm E}(\bY_i^T\bY_i|\bT_i,\tilD_i)&=&\sigma^2(m_i-p_b)+||\rm{E}(\bY_i|\bT_i,\tilD_i)||^2_2,
\ese
in which $\bT_i=\sigma^{-2}\bZ^{bT}_i\bY_i$,
$\bP_{\bZ^b_i}=\bZ^b_i(\bZ^{bT}_i\bZ^b_i)^{-1}\bZ^{bT}_i$, and $\bzeta(\tilD_i;\btheta)=\bZ^a_i\bbeta  + \alpha(\bS_i - X_i\bone_{m_i})$. In Appendix~\ref{sec:calcs_cens}, we show that $\bS_{\rm{eff}}^*(\bY_i,\widehat{\tilD}_i;\btheta)=$
\bse
\label{eqn:cens_eff_score_vec}
\sigma^{-4}\left[
\begin{matrix}
\sigma^2\bZ^{aT}_i(\bI_{m_i}-\bP_{(\bone_{m_i},\bZ^b_i)})\{\bY_i-\bzeta(\widehat{\tilD}_i;\btheta)\} \\
\sigma^2(\bs_i-\widehat{X}_i\bone_{m_i})^T(\bI_{m_i}-\bP_{(\bone_{m_i},\bZ^b_i)})\{\bY_i-\bzeta(\widehat{\tilD}_i;\btheta)\} \\
\frac{1}{2}\{\bY_i^T\bY_i-{\rm E}(\bY_i^T\bY_i|\bT_i^{\rm aug},\widehat{\tilD}_i)\}-\bzeta^T(\widehat{\tilD}_i;\btheta)\{\bY_i-{\rm E}(\bY_i|\bT_i^{\rm aug},\widehat{\tilD}_i)\}
\end{matrix}
\right],
\ese

where
\bse
{\rm E}(\bY_i|\bT_i^{\rm aug},\widehat{\tilD}_i)&=&\bP_{(\bone_{m_i},\bZ^b_i)}\bY_i+(\bI_{m_i}-\bP_{(\bone_{m_i},\bZ^b_i)})\bzeta(\widehat{\tilD}_i;\btheta)\\
{\rm E}(\bY_i^T\bY_i|\bT_i^{\rm aug},\widehat{\tilD}_i)&=&\sigma^2(m_i-p_b)+||\rm{E}(\bY_i|\bT_i^{\rm aug},\widehat{\tilD}_i)||^2_2,
\ese
in which $\bT_i^{\rm aug}=\sigma^{-2}(\bone_{m_i},\bZ^b_i)^T\bY_i$, 
$\bP_{(\bone_{m_i},\bZ^b_i)}=(\bone_{m_i},\bZ^b_i)((\bone_{m_i},\bZ^b_i)^T(\bone_{m_i},\bZ^b_i))^{-1}(\bone_{m_i},\bZ^b_i)^T$,
and $\bzeta(\widehat{\tilD}_i;\btheta)=\bZ^a_i\bbeta  + \alpha(\bS_i - \widehat{X}_i\bone_{m_i})$.

The construction of $S_{\rm eff}(\bY_i,\bD_i;\btheta)$ and $S_{\rm eff}^*(\bY_i,\wh\bD_i;\btheta)$ appears complex, but it involves computationally simple matrix algebra and, perhaps more crucially, completely avoids all terms involving both the random effects $\bb$ and the imputation error $U$. We are able to achieve this result by leveraging the properties of multivariate normal distributions and linear projection operators to show that all terms containing either $\bb$ or $U$ drop out of the efficient score vectors (Appendix~\ref{sec:calcs}). As a result, ACE imputation requires that we neither propose forms for ($f_{\bb}, f_U$) nor estimate these unknown distributions. Therefore, the proposed method is unlike traditional maximum likelihood estimation, which requires that we assume specific forms for ($f_{\bb}, f_U$), such as a multivariate normal distribution. Although maximum likelihood estimators will be more efficient than semiparametric or nonparametric methods when these distributional assumptions are correct, they will suffer from bias when these assumptions are incorrect. The proposed method is also unlike traditional semiparametric estimators, which require positing the nuisance distributions; although these methods produce consistent estimators regardless of whether the posited distributions are correct (an advantage over traditional maximum likelihood estimation), they suffer efficiency losses when the posited distributions are incorrect  \citep{garcia2016optimal}. Since our method does not require positing the nuisance distributions ($f_{\bb}, f_U$), $\ACEest$ achieves both consistency and optimal efficiency without any prior knowledge about the nuisance distributions.

% TEXT FROM TANYA (copied from her 2016 paper):
% This also differs from other semiparametric estimators \citep{TsiatisMa2004,MaCarroll2006,MaGenton2010} where the estimator is constructed under posited working models for the unknown distributions which can affect efficiency
% .  
% For example, in most situations, when one is lucky enough to obtain a consistent
% estimator using a working model for the unknown distribution, the efficiency relies on the
% property of the working model. A classical scenario is when the
% working model happens to be the truth, then optimal efficiency is
% obtainable. Otherwise, only consistency is guaranteed. 
% However, using our approach, the proposed estimator %simultaneously
%  achieves 
% consistency and optimal efficiency without the
% knowledge of the true random effect distribution or the attempt to
% estimate it. Therefore, the estimator is robust to misspecification of the unknown $f_{\bX,R}(\bx_i,r_i)$.

\section{Properties of our Novel Estimator}
\label{sec:ee-properties}

\subsection{Identifiability}
\label{sec:ident}
We now discuss properties of the estimator, $\ACEest$, that solves Equation~\eqref{eqn:total_estimating_equation}. We begin with the identifiability of $\ACEest$:
\begin{Th}
\label{theorem:ident}
%Let $\bS_{\rm{eff}}(\bY_i,\tilD_i;\btheta)$ and $\bS_{\rm{eff}}^*(\bY_i,\widehat{\tilD}_i;\btheta)$ be the efficient score vectors defined in Equations~\eqref{eqn:uncens_eff_score_vec} and \eqref{eqn:cens_eff_score_vec}, respectively, where $\tilD_i=(\bZ^a_i, \bs_i, X_i, \bZ^b_i)$ and $\widehat{\tilD}_i=(\bZ^a_i, \bs_i, \widehat{X}_i, \bZ^b_i)$. 
Let $\bone_n$ denote an $n \times 1$ vector of ones and $\bI_n$ an $n \times x$ identity matrix, and define $\bP_{\bM}=\bM(\bM^T\bM)^{-1}\bM^T$ for a matrix $\bM$. If the matrix
%\bse
$\bN=\sum_{i=1}^n%&&
\Big\{\Delta_i (\bZ^a_i,\bs_i-X_i\bone_{m_i})^T(\bI_m - \bP_{\bZ^b_i})(\bZ^a_i,\bs_i-X_i\bone_{m_i}) + %\\
%&&
(1 - \Delta_i)(\bZ^a_i,\bs_i-X_i\bone_{m_i})^T(\bI_m - \bP_{\bZ^{b*}_i})(\bZ^a_i,\bs_i-X_i\bone_{m_i})\Big\}
$
%\ese
is non-singular (i.e., if $\bN$ is invertible), then $\ACEest$ is identifiable.
\end{Th}
We prove Theorem~\ref{theorem:ident} in Appendix~\ref{sec:ident_proof}. The closed form of $\bN$ in Theorem~\ref{theorem:ident} makes it easy to check for identifiability for a given dataset: we can compute $\bN$ and if it is invertible, we know we can uniquely estimate $\ACEest$ in the presence of the nuisance distributions ($f_{\bb}$, $f_U$). 

\subsection{Consistency}
\label{sec:consistency}

Next, to establish that $\ACEest$ is a consistent estimator for $\btheta_0$, we must first show that the estimating functions in Equation~\eqref{eqn:total_estimating_equation} are unbiased (i.e., that the estimating functions have mean $\bzero$). To this end, we first highlight the simple assumptions that lead to unbiased estimating functions. 

\begin{proposition}
\label{prop:mean_zero}
Consider the following conditions:
\begin{enumerate}
    \item[(C1)] $\bY_i$ is conditionally independent of $C_i$ given $\tilD_i$.
    \item[(C2)] $\bY_i$ is conditionally independent of $C_i$ given $\widehat{\tilD}_i$.
\end{enumerate}
%Next, consider the efficient score vectors $\bS_{\rm{eff}}(\bY_i,\tilD_i;\btheta)$ and $\bS_{\rm{eff}}^*(\bY_i,\widehat{\tilD}_i)$ defined in Equations~\eqref{eqn:uncens_eff_score_vec} and \eqref{eqn:cens_eff_score_vec}, respectively, and let $\Delta_i = I(X_i \leq C_i)$ be the censoring indicator.
%\begin{enumerate}
    %\item[(A)] 
If (C1) holds, then $\E\{\Delta_i\bS_{\rm{eff}}(\bY_i,\tilD_i;\btheta)\} = \bzero$.
    %\item[(B)] 
If (C2) holds, then  $\E\{(1-\Delta_i)\bS_{\rm{eff}}^*(\bY_i,\widehat{\tilD}_i)\} = \bzero$. %The efficient score vectors $\bS_{\rm{eff}}(\bY_i,\tilD_i;\btheta)$ and $\bS_{\rm{eff}}^*(\bY_i,\widehat{\tilD}_i)$ are defined in Equations~\eqref{eqn:uncens_eff_score_vec} and \eqref{eqn:cens_eff_score_vec}, respectively.
\end{proposition}

We prove Proposition~\ref{prop:mean_zero} in Appendix~\ref{sec:mean_zero_proof}. Recall that $\tilD_i=(\bZ^a_i, \bs_i, X_i, \bZ^b_i)$ and that $\widehat{\tilD}_i=(\bZ^a_i, \bs_i, \widehat{X}_i, \bZ^b_i)$. Given these definitions, condition (C1) is equivalent to the claim that once we have observed all covariates of interest, the censoring variable itself gives us no new information about the outcome. Condition (C2) is similar to condition (C1), except that $\tilD_i=(\bZ^a_i, \bs_i, X_i, \bZ^b_i)$ has been replaced by $\widehat{\tilD}_i=(\bZ^a_i, \bs_i, \widehat{X}_i, \bZ^b_i)$ (i.e., $X_i$ has been replaced by the imputed value $\widehat{X}_i$). Whether these conditional independence assumptions are valid will depend on the model and data at hand.

Given this unbiasedness, we establish the consistency of $\ACEest$ by applying the Inverse Function Theorem for likelihood-type estimators \citep{foutz1977unique}, which requires 1) that our estimating functions are unbiased and 2) the technical conditions outlined in Theorem~\ref{theorem:consistency}:
\begin{Th}
\label{theorem:consistency}
Consider the regularity conditions:
\begin{enumerate}
    \item[(C3)] The domain $\boldsymbol{\Theta}$ of the parameter $\btheta$ is a compact set.
    \item[(C4)] The estimating equation $\sum_i\bS_{\rm{eff}}^{\rm{full}}(\bY_i,\tilD_i,\widehat{\tilD}_i;\btheta)$ and $\E\{\sum_i\bS_{\rm{eff}}^{\rm{full}}(\bY_i,\tilD_i,\widehat{\tilD}_i;\btheta)\}$ are sufficiently smooth functions of $\btheta$ in a neighborhood of $\btheta_0$.
    \item[(C5)] The expectation of our estimating equation, $\E\{\sum_i\bS_{\rm{eff}}^{\rm{full}}(\bY_i,\tilD_i,\widehat{\tilD}_i;\btheta)\}$, has a unique solution and each component of $\E\{{\rm sup}_{\btheta \in \Theta}|\sum_i\bS_{\rm{eff}}^{\rm{full}}(\bY_i,\tilD_i,\widehat{\tilD}_i;\btheta)|\}$ is finite.
\end{enumerate}
If conditions (C1)--(C5) hold. then $\ACEest$ is consistent for $\btheta_0$ (i.e., $\ACEest$ converges in probability to $\btheta_0$).
\end{Th}

The consistency of $\ACEest$ is especially appealing given that it does not rely on correctly specifying the nuisance distributions, ($f_{\bb}$, $f_U$). In fact, one does not even need to specify a working model for either of these distributions, and $\ACEest$ still achieves consistency. This contrasts from existing imputation-based solutions to censored covariates \citep{Wang2012, Bernhardt2014, Wei2018, Atem2019, yu2021quantile}, which rely on correctly specifying the underlying imputation model.

% We have also established several key theoretical results regarding ACE imputation. The original form of the estimating function shown in Appendix\ref{sec:original_score_vectors} suggest 1) that it may not have a closed form and 2) that it may depend on (possibly misspecified) nuisance distributions of the random effects $\bb$ and imputation error $U$. However, after taking the necessary derivatives and leveraging key properties about the conditional distribution of the response vector $\bY$, we show that the estimating function does, indeed, have a closed form and does not depend on nuisance distributions. As a result, we arrive at an estimating equation which is straightforward to implement (see Equations~\eqref{eqn:uncens_eff_score_vec} and \eqref{eqn:cens_eff_score_vec}). 

\subsection{Asymptotic normality}
\label{sec:asymp_norm}

$\ACEest$ is not only consistent but also asymptotically normal, as shown next.
\begin{Th}
\label{theorem:asymp_norm}
%Let $\Delta_i=I(X_i\leq C_i)$, let $\bS_{\rm{eff}}(\bY_i,\tilD_i;\btheta)$ and $\bS_{\rm{eff}}^*(\bY_i,\widehat{\tilD}_i;\btheta)\}$ be the efficient score vectors defined in Equations~\eqref{eqn:uncens_eff_score_vec} and \eqref{eqn:cens_eff_score_vec}, let $\bS_{\rm{eff}}^{\rm{full}}(Y_i,\tilD_i,\widehat{\tilD}_i;\btheta) = \Delta_i\bS_{\rm{eff}}(\bY_i,\tilD_i;\btheta) + (1 - \Delta_i)\bS_{\rm{eff}}^*(\bY_i,\widehat{\tilD}_i;\btheta)$ denote the full estimating function, and let $\ACEest$ denote the estimator of $\btheta_0$ found by solving $\sum_{i=1}^n\bS_{\rm{eff}}^{\rm{full}}(\bY_i,\tilD_i,\widehat{\tilD}_i;\btheta)=\bzero$. 
Consider the regularity conditions:
\begin{enumerate}
    \item[(C6)] The matrix $n^{-1}\sum_{i=1}^n\frac{\partial}{\partial\btheta} \{\bS_{\rm{eff}}^{\rm{full}}(\bY_i,\tilD_i,\widehat{\tilD}_i;\btheta)\}$ converges uniformly, in probability, to a matrix $E[\frac{\partial}{\partial\btheta}\{\bS_{\rm{eff}}^{\rm{full}}(\bY_i,\tilD_i,\widehat{\tilD}_i;\btheta)\}]$ in a neighborhood of $\btheta_0$.
    \item[(C7)] The matrix $E[\frac{\partial}{\partial\btheta}\{\bS_{\rm{eff}}^{\rm{full}}(\bY_i,\tilD_i,\widehat{\tilD}_i;\btheta)\}]$ is a bounded, smooth function of $\btheta$ in a neighborhood $\btheta_0$ and is nonsingular. % i.e., there exists a matrix $\bM$ such that $\bM E[\frac{\partial}{\partial\btheta}\{\bS_{\rm{eff}}^{\rm{full}}(\bY_i,\tilD_i,\widehat{\tilD}_i;\btheta)\}]$ is the identity matrix.
\end{enumerate}
Under conditions (C1)--(C7), % and that $\E\{\bS_{\rm{eff}}^{\rm{full}}(\bY_i,\tilD_i,\widehat{\tilD}_i;\btheta)\}=\bzero$
we have that $n^{1/2}(\ACEest-\btheta_0)$ converges in distribution to $\Normal(0, {\cal V})$, where the matrix ${\cal V} = {\cal A}^{-1} {\cal B} ({\cal A}^{-1})^T$ with 
\bse
{\cal A}&=& \E\left\{\frac{\partial \left\{\bS_{\rm{eff}}^{\rm{full}},\tilD_i,\widehat{\tilD}_i;\btheta_0)\right\}}{\partial \btheta^T}\right\}, \ \ \ 
{\cal B}=  \rm{Var}\{\bS_{\rm{eff}}^{\rm{full}},\tilD_i,\widehat{\tilD}_i;\btheta_0)\}.
\ese
\end{Th}
A proof of Theorem~\ref{theorem:asymp_norm} is provided in Appendix~\ref{sec:asymp_norm_proof}. This result is powerful because if conditions (C1)--(C2) (Proposition~\ref{prop:mean_zero}), (C3)--(C5) (Theorem~\ref{theorem:consistency}), and (C6)--(C7) (Theorem~\ref{theorem:asymp_norm}) all hold, then we can perform asymptotically valid inference about our parameters of interest, $\btheta$. For example, we can construct Wald-type confidence intervals for $\btheta$ and conduct asymptotically valid hypothesis tests about $\btheta$. Moreover, this asymptotic inference can be done without specifying or estimating the nuisance distributions, ($f_{\bb}$, $f_U$). Again, this method is unlike existing methods, which require correctly specifying the imputation model.

\section{Simulation study}
\label{sec:simulation}

We next compare how ACE imputation performs vs. a competitor in simulation studies. The competitor is multiple conditional mean imputation, where we perform conditional mean imputation $M$ times, estimate $\btheta$ by applying REML to the multiple imputed datasets, and pool the $M$ sets of parameter estimates. The core difference between ACE imputation and the competitor is that our method adjusts for errors that occur when the imputation model is incorrect, whereas the competitor does not adjust for these errors. 

% In the first, we impute the censored covariate using a correctly specified imputation model and then estimate a longitudinal model with those imputed covariates using our novel estimating equation (Equation~\eqref{eqn:total_estimating_equation}). The second simulation study is identical, except that we impute the censored covariate using a misspecified imputation model. In both settings, we compare ACE imputation to an existing approach for overcoming censored covariates. In the first simulation study, the existing solution performs moderately worse than ACE imputation, which produces highly accurate parameter estimates and reliable inference (Section~\ref{sec:sim_A}). In the second setting, we see that the existing solution performs much worse as a result of imputation model misspecification, whereas the accuracy and efficiency of ACE imputation are largely unimpeded (Section~\ref{sec:sim_B}). This is a compelling result, since model misspecification occurs, at least to some extent, in many biological settings \citep{haber2020bias}. We begin by describing the data generation mechanisms for our simulations.

\subsection{Data generation}
\label{sec:sim-setup}
%We begin by describing the data generation mechanisms for our simulations.
We simulate data for $n=1000$ subjects with $m=3$ observations each. We simulate times of clinical diagnosis $X_i$ in two ways. In the first simulation setting, we generate $X_i$ according to a Cox model with two distinct covariates; later in this setting, we impute censored values of $X_i$ using a Cox model that includes both of these covariates, so the imputation model is correctly specified. In the second simulation setting, we generate $X_i$ according to a Cox model with linear and quadratic terms for the same covariate; when we impute censored values of $X_i$ in this setting, we omit the quadratic term and hence misspecify the imputation model. The hazard functions to generate these data are:
\be
\lambda(t|\bV_i)&=&\lambda_0(t)\exp(\eta_{11}V_{i1}-\eta_{12}V_{i2})\label{eqn:sim_cox_correct_A},%~\textrm{and}
\\
\lambda(t|\bV_i)&=&\lambda_0(t)\exp(\eta_{21}V_{i1}-\eta_{22} V_{i1}^2)\label{eqn:sim_cox_correct_B},
\ee
corresponding to the correctly specified imputation model and incorrectly specified imputation model settings, respectively. We set the log-hazard ratios $(\eta_{11},\eta_{12})\trans=(1,0.5)\trans$ and $(\eta_{21},\eta_{22})\trans=(1,0.25)\trans$. The covariates $V_{i1}, V_{i2}$ are independent and normally distributed with mean $0$ and variance $1$. To generate $X_i$ so that it follows the Cox models in Equations~\eqref{eqn:sim_cox_correct_A} and~\eqref{eqn:sim_cox_correct_B}, we follow the strategy outlined by \cite{bender2005generating}.%: we generate the random variables $U_i$ from a uniform $[0,1]$ distribution and calculate $X_i=-2\log(U_i)\exp(1V_{i1}-0.5V_{i2})$ (Equation~\eqref{eqn:sim_cox_correct_A}) or $X_i=-2\log(U_i)\exp(1V_{i1}-0.25V_{i1}^2)$ (Equation~\eqref{eqn:sim_cox_correct_B}). % In each of these calculations, the leading $2$ is included so that $X_i$ follows an exponential distribution with rate parameter $1/2$.

We generate $Y_{ij}$ according to the following linear mixed model:
\bse
&&Y_{ij} =
\beta Z^a_{ij} + \alpha(s_{ij}-X_i) + b_iZ_{ij} + \epsilon_{ij}, \ \ \
b_i\sim f_b, \ \ \ \epsilon_{ij}\sim\Normal(0,\sigma^2), \nonumber
\ese
where $\btheta=(\beta,\alpha,\sigma^2)\trans=(1,1,1)\trans$. We generate $Z^a_{ij}\sim\Normal(0,1)$, $Z_{ij}\sim\Normal(5,1)$,  
%BECAUSE OUR EE NEEDS TO USE $(\sum_j Z_{ij}^2)^{-1}$, WHICH BLOWS UP WHEN THE $Z_{ij}$ ARE NEAR 0., 
$b_i\sim\Normal(0,1)$, and $s_{ij}=j-1$. Having $s_{ij}=j-1$ creates data where subsequent observations for the same subject are evenly spaced by $1$ unit of time and observations begin at time = $0$; this choice of $s_{ij}$ replicates the PREDICT-HD data, where the average time between subsequent observations is $\approx 1.18$ years. Lastly, $X_i$ was censored by a random right-censoring variable $C_i$ that we generated from an exponential distribution with varied rate parameter $\lambda_C$. Having $\lambda_C=0.125,0.5$, and $2$ led to light ($\approx 25\%$), medium ($\approx 50\%$), and heavy ($\approx 75\%$) censoring, respectively. %We calculate the observed times $W_i=\min(X_i, C_i)$ and event indicators $\Delta_i=I(X_i<C_i)$ for all $i$ so that the time of clinical diagnosis $X_i$ is censored for some subjects in our data.

% Moreover, by varying $\lambda_C$, we are able to vary the censoring rate; with both data generation models (Equations~\eqref{eqn:sim_cox_correct_A} and~\eqref{eqn:sim_cox_correct_B}, we achieve light ($\approx 21\%$ censoring), medium ($\approx 40\%$ censoring), and heavy ($\approx 59\%$ censoring) censoring rates with $\lambda_C=0.1,0.3,$ and $0.8$, respectively. 

\subsection{Model evaluation methods}
\label{sec:model_eval}

For any $X_i$ that is censored, we impute the censored values using conditional mean imputation (Section~\ref{sec:cmi}). First, we impute using a model that is correctly specified, i.e., we include linear terms for both $V_{i1}$ and $V_{i2}$ in accordance with Equation~\eqref{eqn:sim_cox_correct_A}. In the second setting, however, we misspecify the imputation model by omitting the quadratic term $V_{i}^2$ from Equation~\eqref{eqn:sim_cox_correct_B} and fit the incorrect imputation model:
%\bse
$\lambda^*(t|V_{i1})=\lambda_0(t)\exp(\eta_{21}^*V_{i1})$,
%\ese
which is misspecified given the true data generation mechanism in Equation~\eqref{eqn:sim_cox_correct_B}. 

We simulated 1000 datasets and, for each, we estimate $\btheta$ using the following methods: % (see Figure~\ref{fig:sim-flowchart}):
\begin{enumerate}
    \item \emph{Oracle estimator}: We apply REML to the full, uncensored dataset. The result from this method is what we would obtain had the data never been censored. REML is a standard procedure for estimating linear mixed models without censored data \citep{ALA}. Of course, this approach is not feasible in practice when times of clinical diagnosis have been censored; hence, we consider these estimates to be the ``gold standard'' or ``Oracle'' in our simulations. 
    % \item \emph{Conditional Mean Imputation (CMI)}: We replace all censored values of $X_i$ with their conditional means $\widehat{X}_i$, computed using conditional mean imputation. We then apply REML to this imputed dataset; we consider these estimates to be the ``competing solution'' to covariate censoring in a longitudinal setting. We expect this competing method to be unbiased when the imputation model is correctly specified, but biased when the imputation model is misspecified.
    \item \emph{Multiple Conditional Mean Imputation (MCMI):} We also apply a multiple imputation procedure that incorporates conditional mean imputation. We repeat the imputation procedure described in Section~\ref{sec:cmi} above $M=15$ times, but with $\widehat{\boldsymbol{\eta}}$ --- the vector of estimated log-hazard ratios in the imputation model --- instead drawn from  $\Normal\{\widehat{\boldsymbol{\eta}}, {\rm Cov}(\widehat{\boldsymbol{\eta}})\}$ \citep{cole2006multiple}. We then pool these $M$ sets of parameter estimates using Rubin's rules for multiple imputation. We expect this method to be unbiased when the imputation model is correctly specified, but biased when the imputation model is misspecified.
    \item \emph{ACE imputation (ACE)}: We apply our proposed method, which we expect to be unbiased regardless of whether the imputation model has been correctly specified.
\end{enumerate}

% \begin{figure}
%     \centering

% \begin{tikzpicture}[node distance=2cm]
% \node (start) [startstop] {Simulate Data};
% \node (in1) [io, below of=start] {Censor T?};
% \node (pro1a) [method1, below of=in1] {Estimate $\btheta$ with Restricted MLE};
% \node (pro1b) [process, right of=in1, xshift=3.5cm] {Impute T with Conditional Mean Imputation};
% \node (pro2a) [method2, below of=pro1b] {Estimate $\btheta$ with Res\ttricted MLE};
% \node (pro2b) [method3, right of=pro2a, xshift=3.5cm] {Estimate $\btheta$ with ACE imputation};

% \draw [arrow] (start) -- (in1);
% \draw [arrow] (in1) -- node[anchor=east] {No} (pro1a);
% \draw [arrow] (in1) -- node[anchor=south] {Yes}(pro1b);
% \draw [arrow] (pro1b) -- (pro2a);
% \draw [arrow] (pro1b) -- (pro2b);
% \end{tikzpicture}
    
%     \caption{Simulation flowchart}
%     \label{fig:sim-flowchart}
% \end{figure}

To evaluate the performance of these methods, we calculate the empirical bias, the empirical mean of the standard error estimates, the empirical standard deviation of the parameter estimates, and the empirical mean of the squared biases. We also report the observed coverage of the Wald-type confidence intervals with nominal 95\% coverage. We do not present standard error estimates and coverage probabilities for $\sigma^2$ from REML, since the variability of $\sigma^2$ is not typically of interest with this method. Results under medium and heavy censoring are shown in Table~\ref{table:corr_spec_results} and Table~\ref{table:mis_spec_results}. Results under light censoring are shown in Table~\ref{table:corr_spec_results_light} and Table~\ref{table:mis_spec_results_light} in Appendix~\ref{sec:app_tables}.

We employ both existing and new software to carry out these imputation and estimation procedures in R \citep{R-lang}. We use the  {\tt coxph()} function from the \textit{survival} package to estimate all Cox models \citep{survival-package}; the
{\tt condl\_mean\_impute()} function from the \textit{imputeCensoRd} package to execute conditional mean imputation  \citep{correctingCMI}; and the {\tt lmer()} function from the \textit{lme4} package to carry out REML \citep{lme4}. To implement ACE imputation, we developed an R package, \textit{ACEimpute}; this package contains the {\tt eff\_score\_vector()} function, which constructs the estimating equation shown in Equation~\eqref{eqn:total_estimating_equation}. Lastly, to solve for the root of this estimating equation, we use the {\tt m\_estimate()} function from the \textit{geex} package \citep{Saul2020geex}.

\subsection{A correctly specified imputation model}
\label{sec:sim_A}
ACE imputation yields highly accurate parameter estimates when the imputation model is correctly specified (Table~\ref{table:corr_spec_results}). The average empirical bias from ACE imputation is $< 0.003$ for each parameter under light, medium, and heavy censoring. Since the true parameters are all 1, this finding is equivalent to an average percent bias $< 1\%$. Furthermore, we see that the average standard error estimates capture the true variability of the parameter estimates. Together, these results lead to valid inference about $\alpha$ and $\beta$; specifically, we see that 
%when we use our estimating equation to generate Wald-type confidence intervals with nominal coverage probability = 95\%,
the observed coverage probability of the $95\%$ confidence intervals for $\alpha$ and $\beta$ are between 94\% and 96\% under each censoring rate.
% These coefficients are the target of our inference

Furthermore, ACE imputation outperforms the competitor, MCMI. We expected that the competitor would produce unbiased parameter estimates since conditional mean imputation (followed by ordinary least squares regression, instead of REML) is empirically unbiased in cross-sectional settings \citep{Atem2019}. In our longitudinal simulations, however, MCMI estimates $\alpha$ with $\geq 20\%$ bias, on average. This approach also yields an estimate of $\sigma^2$ with bias $\geq 150\%$ under all three censoring rates. We see that MCMI does yield accurate estimates of $\beta$ (with $\approx 0$ bias on average), but these estimates of $\beta$ are more variable than those produced by ACE imputation. Based on these findings, MCMI leads to seriously biased estimates of $\alpha$ and $\sigma^2$. Fortunately, ACE imputation provides a powerful remedy to this estimation bias, while estimating $\beta$ more efficiently.

% latex table generated in R 4.1.0 by xtable 1.8-4 package
% Thu Nov 17 11:12:55 2022
\begin{table}[!ht]
\centering
\begin{tabular}{lllrrrrr}
  \hline
Censoring & Param. & Method & Bias & SEE & ESE & MSE & CPr \\ 
  \hline
   
   None & $\alpha$ & Oracle & -0.001 & 0.009 & 0.010 & 0.000 & 0.928 \\
   & $\beta$ & Oracle & 0.000 & 0.022 & 0.022 & 0.000 & 0.940 \\ 
   & $\sigma^2$ & Oracle & -0.002 &  & 0.032 & 0.001 &  \\ 
% Light & $\alpha$ & ACE & -0.000 & 0.022 & 0.022 & 0.000 & 0.950 \\ 
%   &  & CMI & 0.197 & 0.022 & 0.132 & 0.056 & 0.046 \\ 
%   &  & MCMI & 0.173 & 0.023 & 0.127 & 0.046 & 0.070 \\ 
%   & $\beta$ & ACE & 0.000 & 0.024 & 0.024 & 0.001 & 0.942 \\ 
%   &  & CMI & 0.002 & 0.036 & 0.037 & 0.001 & 0.957 \\ 
%   &  & MCMI & 0.002 & 0.036 & 0.038 & 0.001 & 0.951 \\ 
%   & $\sigma^2$ & ACE & -0.003 & 0.034 & 0.034 & 0.001 & 0.940 \\ 
%   &  & CMI & 1.819 &  & 2.911 & 11.771 &  \\ 
%   &  & MCMI & 1.940 &  & 2.979 & 12.629 &  \\
  Medium & $\alpha$ & ACE & 0.000 & 0.026 & 0.024 & 0.001 & 0.960 \\ 
  ($\approx 50\%$) % &  & CMI & 0.312 & 0.033 & 0.107 & 0.109 & 0.000 \\ 
   &  & MCMI & 0.312 & 0.033 & 0.107 & 0.109 & 0.000 \\
   & $\beta$ & ACE & 0.001 & 0.026 & 0.025 & 0.001 & 0.958 \\ 
   % &  & CMI & 0.001 & 0.040 & 0.040 & 0.002 & 0.966 \\ 
   &  & MCMI & 0.001 & 0.040 & 0.040 & 0.002 & 0.966 \\
   & $\sigma^2$ & ACE & -0.000 & 0.036 & 0.037 & 0.001 & 0.940 \\ 
   % &  & CMI & 2.497 &  & 2.946 & 14.909 &  \\ 
   &  & MCMI & 2.498 &  & 2.947 & 14.914 &  \\
  Heavy & $\alpha$ & ACE & 0.001 & 0.028 & 0.029 & 0.001 & 0.947 \\ 
  ($\approx 75\%$) % &  & CMI & 0.206 & 0.044 & 0.072 & 0.048 & 0.015 \\ 
   &  & MCMI & 0.206 & 0.044 & 0.072 & 0.048 & 0.014 \\
   & $\beta$ & ACE & 0.001 & 0.028 & 0.028 & 0.001 & 0.952 \\ 
   % &  & CMI & -0.001 & 0.045 & 0.046 & 0.002 & 0.951 \\ 
   &  & MCMI & -0.001 & 0.045 & 0.046 & 0.002 & 0.950 \\
   & $\sigma^2$ & ACE & -0.003 & 0.040 & 0.040 & 0.002 & 0.940 \\ 
   % &  & CMI & 3.356 &  & 3.426 & 22.987 &  \\ 
   &  & MCMI & 3.356 &  & 3.426 & 22.989 &  \\
   \hline
\end{tabular}
\caption{Simulation results when the imputation model is correctly specified. Results are shown for three methods: ACE imputation, % CMI (conditional mean imputation), 
MCMI (multiple conditional mean imputation), and the Oracle estimator. We present empirical bias, the average standard error estimate (SEE), the empirical standard error (ESE), the average mean squared errors (MSE), and the observed coverage probability (CPr) of the Wald-type confidence intervals with nominal 95\% coverage.}
\label{table:corr_spec_results}
\end{table}

\subsection{A misspecified imputation model}
\label{sec:sim_B}
Table~\ref{table:mis_spec_results} shows the results when we omit the quadratic term from (and hence, misspecify) the imputation model. We see that the competitor, MCMI, performs much worse under this setting than when the imputation model is correct. Although this approach still estimates $\beta$ quite accurately, it now estimates $\alpha$ with $> 100\%$ bias under all three censoring rates and drastically overestimates $\sigma^2$, with an average bias $>> 100\%$. This bias is to be expected; when we misspecify our imputation model, we should anticipate that the conditional means ${\rm E}(X_i|X_i>C_i,V_{i1})$, with which we replace censored $X_i$, should be further away from the true $X_i$. In contrast, we see that imputation model misspecification does not hinder the performance of ACE imputation, which performs very similarly in both settings considered. ACE imputation produces accurate parameter estimates and reliable inference even though we have omitted the quadratic term from our imputation model.
% Since model misspecification can easily occur in biological settings \citep{haber2020bias}, 
ACE imputation thus allows us to consistently estimate the parameters of a linear mixed model with a censored covariate, even when we do not know the correct imputation model \textit{a priori}, which can easily occur in biological settings \citep{haber2020bias}. 

% Indeed, the average empirical bias from ACE is $< 0.004$ for each parameter, regardless of the censoring rate; again, this implies an average percent bias $< 1\%$. As in the Correctly Specified setting, we also observe the standard error estimates accurately reflect the true variability of the parameter estimates. This, again combined with the unbiased estimates, leads to valid inference about $\alpha$ and $\beta$: the observed coverage probability of the 95\% confidence intervals for $\alpha$ and $\beta$ are between 95\% and 97.0\% under each censoring rate.

% This demonstrates that conditional mean imputation \textit{on its own} is susceptible to imputation model misspecification, whereas misspecification does not reduce the accuracy or inferential validity of our proposed estimation procedure.

\begin{table}[!ht]
\centering
\begin{tabular}{lllrrrrr}
  \hline
Censoring & Param. & Method & Bias & SEE & ESE & MSE & Coverage \\ 
  \hline
   None & $\alpha$ & Oracle & -0.000 & 0.003 & 0.003 & 0.000 & 0.954 \\
   & $\beta$ & Oracle & 0.001 & 0.022 & 0.022 & 0.000 & 0.953 \\ 
   & $\sigma^2$ & Oracle & -0.001 &  & 0.032 & 0.001 &  \\ 
%   Light & $\alpha$ & ACE & 0.000 & 0.022 & 0.022 & 0.000 & 0.946 \\ 
%   &  & CMI & 1.874 & 0.173 & 10.452 & 112.645 & 0.000 \\ 
%   &  & MCMI & 1.872 & 0.175 & 10.429 & 112.154 & 0.000 \\ 
%   & $\beta$ & ACE & 0.001 & 0.024 & 0.024 & 0.001 & 0.948 \\ 
%   &  & CMI & -0.071 & 0.316 & 1.287 & 1.660 & 0.953 \\ 
%   &  & MCMI & -0.071 & 0.316 & 1.289 & 1.665 & 0.953 \\ 
%   & $\sigma^2$ & ACE & -0.002 & 0.034 & 0.035 & 0.001 & 0.944 \\ 
%   &  & CMI & 10575.377 &  & 299437.561 & 89685028812.978 &  \\ 
%   &  & MCMI & 10577.654 &  & 299510.122 & 89728493594.018 &  \\ 
  Medium & $\alpha$ & ACE & 0.000 & 0.026 & 0.026 & 0.001 & 0.948 \\ 
  ($\approx 50\%$) % &  & CMI & 1.592 & 0.263 & 6.584 & 45.837 & 0.005 \\ 
   &  & MCMI & 1.591 & 0.264 & 6.569 & 45.639 & 0.005 \\
   & $\beta$ & ACE & 0.001 & 0.026 & 0.026 & 0.001 & 0.949 \\ 
   % &  & CMI & -0.069 & 0.327 & 1.136 & 1.294 & 0.955 \\ 
   &  & MCMI & -0.069 & 0.327 & 1.137 & 1.296 & 0.955 \\
   & $\sigma^2$ & ACE & -0.002 & 0.037 & 0.038 & 0.001 & 0.943 \\ 
   % &  & CMI & $\geq 100$ &  & $\geq 100$ & $\geq 100$ &  \\ 
   &  & MCMI & $\geq 100$ &  & $\geq 100$ & $\geq 100$ &  \\
  Heavy & $\alpha$ & ACE & 0.000 & 0.029 & 0.029 & 0.001 & 0.946 \\ 
  ($\approx 75\%$) % &  & CMI & 0.718 & 0.324 & 3.100 & 10.113 & 0.189 \\ 
   &  & MCMI & 0.716 & 0.326 & 3.085 & 10.018 & 0.188 \\
   & $\beta$ & ACE & 0.000 & 0.029 & 0.029 & 0.001 & 0.948 \\ 
   % &  & CMI & -0.073 & 0.332 & 1.262 & 1.596 & 0.949 \\ 
   &  & MCMI & -0.073 & 0.332 & 1.263 & 1.598 & 0.949 \\ 
   & $\sigma^2$ & ACE & -0.001 & 0.041 & 0.040 & 0.002 & 0.951 \\ 
   % &  & CMI & $\geq 100$ &  & $\geq 100$ & $\geq 100$ &  \\ 
   &  & MCMI & $\geq 100$ &  & $\geq 100$ & $\geq 100$ &  \\ 
   \hline
\end{tabular}
\caption{Simulation results when the imputation model is misspecified. Results are shown for three methods: ACE imputation, % CMI (conditional mean imputation), 
MCMI (multiple conditional mean imputation), and the Oracle estimator. We present empirical bias, the average standard error estimate (SEE), the empirical standard error (ESE), the average mean squared errors (MSE), and the observed coverage probability (CPr) of the Wald-type confidence intervals with nominal 95\% coverage.}
\label{table:mis_spec_results}
\end{table}

\section{Modeling the progression of Huntington disease}
\label{sec:real_data_analysis}

Huntington disease is a fatal neurodegenerative disorder that causes impairment across motor, cognitive, and psychiatric domains. The disease is caused by repeated cytosine-adenine-guanine (CAG) mutations in the huntingtin gene and people with more than 36 CAG repeats are guaranteed to develop Huntington disease \citep{mccolgan2018huntington}. This creates a unique opportunity for clinicians studying the disease; with genetic testing, it is possible to recruit subjects who are guaranteed to develop Huntington disease (``gene mutation carriers'') and test therapies aimed at slowing its progression.

A key consideration when planning clinical trials of Huntington disease is selecting outcomes that change quickly enough over time, so that analysts can more easily detect differences between the treatment and placebo groups. To facilitate this selection, analysts have sought to compare possible outcomes in terms of how quickly they change in gene mutation carriers \citep{langbehn2020clinical, paulsen2014clinical}. To quantify the speed at which these outcomes change, analysts estimate longitudinal models for these outcomes using data from untreated subjects; these ``disease progression models'' then give a metric for how quickly the possible outcomes progress over time in untreated subjects. Hence, by comparing these estimated disease progression models, we can easily compare how quickly these outcomes progress.

To contribute to this ongoing effort, we analyze data from PREDICT-HD, a longitudinal study of gene mutation carriers who were recruited prior to diagnosis \citep{paulsen2008detection}. Specifically, we assess seven potential outcomes that quantify Huntington disease symptoms: (1) total motor score (TMS); (2) SDMT score; (3--5) scores on the Stroop word, color, and interference tests; (6) total functional capacity (TFC); and (7) composite unified Huntington disease rating scale (cUHDRS). TMS assesses a subject's motor impairment. The SDMT and Stroop tests assess the degree of cognitive impairment. TFC assesses a subject's functional ability to perform daily tasks. The cUHDRS is a linear combination of TFC, TMS, SDMT score, and the Stroop word test score \citep{schobel2017motor}.

For each outcome $Y_{ij}$, we fit the following linear mixed effects model:
\be
\label{eqn:real-data-model}
&&Y_{ij} = \beta_0 + \bbeta^\textrm{T} \bZ_{ij} + \alpha(s_{ij} - X_i) + b_i + \epsilon_{ij}, %\ \ \  b_i\sim f_{b}, \ \ \ \epsilon_{ij}\sim\Normal(0,\sigma^2),
\ee
with fixed intercept $\beta_0$, fixed slopes ($\bbeta^\textrm{T}$, $\alpha$), and random intercept $b_i$. The random intercepts $b_i$ are assumed to follow the distribution $f_b$, which is left unspecified, and the random errors $\epsilon_{ij}$ are assumed to be normally distributed with mean $0$ and variance $\sigma^2$. For each subject $i$, the additional covariates $\bZ_{ij}=({\rm AGE}_i,{\rm SEX}_i,{\rm EDUCATION}_i,{\rm CAG}_i)$ are included, all measured at baseline. ${\rm AGE}_i$ is age at baseline, ${\rm SEX}_i$ is $1$ if subject $i$ is male and $0$ if subject $i$ is female, ${\rm EDUCATION}_i$ is years of education, and ${\rm CAG}_i$ is number of CAG repeats \citep{paulsen2014clinical}.
The variables $s_{ij}$ and $X_i$ denote time in the study (in years) and time of clinical diagnosis (in years), respectively, such that $s_{ij} - X_i$ denotes time to clinical diagnosis. As discussed in Section~\ref{sec:intro}, we choose to use time to clinical diagnosis ($s_{ij} - X_i$) instead of just time in the study ($s_{ij}$) to account for the fact that subjects enter the study at different levels of disease progression. %Moreover, we leave the distribution of the random intercept $b_i$ ($f_b(\cdot)$) unspecified to avoid possible misspecification. %(see Section \ref{sec:Hilbert-space-stuff} for details).

To compare how the potential outcomes progress in gene mutation carriers, we compare the estimated slopes on time to clinical diagnosis (i.e., $\hat{\alpha}$) when the model in Equation~\eqref{eqn:real-data-model} is fit for each outcome (cUHDRS, TMS, SDMT score, TFC, and the Stroop test scores). The outcomes are ranked based on their ``standardized slopes,'' calculated as the absolute values of the slopes divided by their estimated standard errors  (i.e., $|\hat{\alpha}|/{\rm SE}(\hat{\alpha})$).

Prior to analysis, we apply exclusion criteria similar to those employed by \cite{long2017validation}. We require subjects to be a Huntington disease carrier (i.e., to have $\geq$ 36 CAG repeats) and not yet 
be clinically diagnosed at study entry (i.e., to have a diagnostic confidence level $\leq 3$ at their first visit). We also filter to include only those visits that have complete data for all outcomes and covariates $Z_{ij}$ of interest, which removes only 25 subjects (2\%) and 260 visits (4\%). %; since the percent of missing-ness with respect to these variables is so low, we feel that complete case analysis in this respect is appropriate. 
With these criteria, we arrive at an analytic dataset containing 1,102 unique subjects with 5,612 total observations. Descriptive statistics are provided in Table~\ref{table:TableOne} in Appendix~\ref{sec:app_tables}.

\subsection{Imputing censored times of clinical diagnosis}
\label{sec:real_data_cmi}
Of the 1,102 subjects, 244 (22.1\%) were clinically diagnosed during the course of study, and 858 (77.9\%) were not (i.e., their times of diagnosis were right-censored). We imputed all censored times of diagnosis using the Cox-based conditional mean imputation (Section~\ref{sec:cmi}).

We use CAG-Age product, TMS, and SDMT score as our set of Cox model covariates. The CAG-Age Product is defined as $X_{i, \rm age} \times (X_{i, \rm CAG} - 34)$ \citep{zhang2011indexing} and conveys the ``burden of disease'' that has accumulated over a subject's life. \cite{long2017validation} show that these three variables predict time of clinical diagnosis better than competing sets of covariates, and we believe that these variables are also an intuitive choice for predicting time of clinical diagnosis. It has been well established that the number of CAG repeats is negatively associated with time of clinical diagnosis \citep{long2017validation}. In addition, TMS and SDMT score measure a subject's motor and cognitive capacities, respectively, which are both known to decrease as a subject approaches clinical diagnosis. With these covariates, we fit a Cox model with hazard function
%\bse
$\lambda(t|\bV_i)=\lambda_0(t)\exp(\bgamma\bV_i)$
%\ese 
where covariates $\bV_i$ include the visit 1 values of TMS, SDMT score, and CAG-Age Product for subject $i$ and $\bgamma = (\gamma_1,\gamma_2,\gamma_3)\trans$ are the corresponding log-hazard ratios. Before fitting this model, we center and scale the covariates using their sample means and standard deviations (given in Table~\ref{table:TableOne} in Appendix~\ref{sec:app_tables}) to reduce possible collinearity.
% Using the function {\tt coxph} from the package \textit{survival} \citep{survival-package}, we obtain log-hazard ratio estimates $\hat{\bgamma}=(\hat{\gamma}_1,\hat{\gamma}_2,\hat{\gamma}_3) = (1.34, 0.65, 1.90)$ with corresponding 95\% confidence intervals $(1.22, 1.47)$, $(0.56, 0.76)$, and $(1.68, 2.15)$. 
% Higher values of CAG-Age Product and TMS as well as lower scores on the SDMT are associated with increased hazard of experiencing time of clinical diagnosis.
With this Cox model estimate, we implement conditional mean imputation to replace values of $X_i$ with $\widehat{X}_i=\Delta_iX_i+(1-\Delta_i){\rm E}(X_i|X_i>C_i,\bV_i)$.

\subsection{Ranking measures of Huntington disease impairment}
\label{sec:ranking_outcomes}

Using this imputed dataset, we estimate the linear mixed effects model in Equation~\eqref{eqn:real-data-model} for each of the seven outcomes. We compute how quickly each potential outcome changed over time using the scaled slope $|\hat{\alpha}|/ {\rm SE}(\hat{\alpha})$ and rank them. We carry out this ranking procedure once for each of three methods: i) complete case analysis (CCA), with REML applied to uncensored subjects only, ii) multiple conditional mean imputation (described in Section~\ref{sec:model_eval}), and iii) ACE imputation.
% \begin{enumerate}
    % \item \emph{Complete Case Analysis (CCA)}: We apply REML to uncensored subjects only (i.e., complete case analysis). Because the censoring rate is so high in our dataset ($\approx 78\%$), this method ``throws away'' most of the available data and hence, we expect results from this approach to be highly variable and possibly biased.
    % \item \emph{Multiple Imputation (MCMI):} To better utilize the available data, we apply REML estimation to datasets that have been ``filled in'' using conditional mean imputation (see Section~\ref{sec:model_eval} for more detail on the multiple imputation framework). We consider this the ``competing solution'' to covariate censoring in a longitudinal setting, and we expect this method to be more efficient than complete case analysis since it uses all available data. However, as we demonstrated in Section~\ref{sec:simulation}, this approach incurs bias even when the imputation model is correctly specified. 
    % \item \emph{ACE imputation}: Hence, to reduce this bias, we apply ACE imputation to the imputed dataset. We have shown in simulation studies that this method reduces estimation bias on $\alpha$ by actively adjusting for the error incurred when we impute (Section~\ref{sec:berkson}), even when the imputation model has been misspecified.
% \end{enumerate}
Table~\ref{table:rank_slopes} shows the results of these ranking procedures. All three methods agree that the cUHDRS shows the quickest progression. This finding is in contrast to that of \cite{langbehn2020clinical}, who 
% also attempt to pinpoint a meaningful outcome for clinical trials of Huntington disease by ranking possible outcomes; their results 
show that TMS changes most quickly among the outcomes analyzed and is therefore the preferred outcome. 

Whereas TMS measures only motor impairment, the cUHDRS is a linear combination of TFC, TMS, SDMT score, and Stroop word test score; hence, the cUHDRS is a more comprehensive measure of the impairment caused by Huntington disease. Identifying a therapy that improves subjects' cUHDRS could, therefore, more holistically improve quality of life. 
This holistic improvement is a main reason why the cUHDRS is being advocated for as a primary outcome for clinical trials targeting Huntington disease in subjects who have already been diagnosed. Having analyzed data exclusively from patients who were undiagnosed at baseline, our results suggest that the cUHDRS could also be chosen as an easily detectable outcome for patients recruited in this ``pre-manifest'' stage.

% Although this warrants further investigation, there is a preference for cUHDRS over TMS because TMS captures just one aspect of Huntington disease, which is known to present with a triad of symptoms. It has been suggested that cUHDRS is a meaningful outcome for manifest trials; our work suggests that it could also be used for pre-manifest trials.

It should be noted that since the cUHDRS is a linear combination of four other measures, it could be more prone to missing-ness for a given dataset. Given this possible concern, investigators may want to measure impairment due to Huntington disease using an outcome which measures only one symptom (e.g., SDMT score, TMS). According to our results, the best outcome that is a measure of only one symptom is SDMT score. In fact, it has been established that cognitive impairment shows up earlier than motor impairment \citep{paulsen2008detection}. Our analysis aligns with this previous finding, whereas the other methods point to motor impairment instead.

% We also see that complete case analysis and conditional mean imputation produce nearly the same ranking of outcomes. These two methods agree that total motor score shows the steepest slope among individual domains (i.e., excluding cUHDRS, which combines several domains). In contrast, the results from ACE imputation suggest that the symbol digits modality test, a measure of cognitive ability, shows the steepest slope among individual domains. Based on the estimates from ACE imputation, we also see that the slope on the symbol digit modality test is nearly as steep as that on the cUHDRS (proportion $=0.91$). This suggests that changes in cognitive ability may produce more robust signals than changes in other individual domains of Huntington disease impairment.

% latex table generated in R 4.2.0 by xtable 1.8-4 package
% Fri Aug  5 16:07:52 2022
\begin{table}[hbt!]
\footnotesize
\centering
\begin{tabular}{llrrrrr}
  \hline
Method & Outcome & Rank & Estimate & SE & Scaled Slope & Prop. to Max. \\ 
  \hline
CCA & cUHDRS & 1 & -0.506 & 0.011 & 44.179 & 1.000 \\ 
  & TMS & 2 & 2.304 & 0.062 & 37.292 & 0.844 \\ 
  & Stroop Word & 3 & -2.106 & 0.070 & 30.118 & 0.682 \\ 
  & SDMT & 4 & -1.258 & 0.043 & 29.205 & 0.661 \\ 
  & Stroop Color & 5 & -1.608 & 0.060 & 26.629 & 0.603 \\ 
  & TFC & 6 & -0.253 & 0.010 & 24.104 & 0.546 \\ 
  & Stroop Interference & 7 & -0.762 & 0.044 & 17.419 & 0.394 \\ 
  \rowcolor[gray]{0.75}ACE & cUHDRS & 1 & -0.253 & 0.013 & 20.064 & 1.000 \\ 
  \rowcolor[gray]{0.75}& SDMT & 2 & -0.757 & 0.041 & 18.303 & 0.912 \\ 
  \rowcolor[gray]{0.75}& Stroop Word & 3 & -1.177 & 0.070 & 16.883 & 0.841 \\ 
  \rowcolor[gray]{0.75}& TMS & 4 & 0.968 & 0.064 & 15.026 & 0.749 \\ 
  \rowcolor[gray]{0.75}& Stroop Color & 5 & -0.802 & 0.059 & 13.532 & 0.674 \\ 
  \rowcolor[gray]{0.75}& TFC & 6 & -0.102 & 0.009 & 11.466 & 0.571 \\ 
  \rowcolor[gray]{0.75}& Stroop Interference & 7 & -0.314 & 0.037 & 8.489 & 0.423 \\ 

  MCMI & cUHDRS & 1 & -0.250 & 0.005 & 45.714 & 1.000 \\ 
  & TMS & 2 & 1.068 & 0.024 & 44.250 & 0.968 \\
  
  & Stroop Word & 3 & -1.170 & 0.039 & 29.876 & 0.654 \\ 
  & SDMT & 4 & -0.739 & 0.026 & 28.226 & 0.617 \\ 
  & TFC & 5 & -0.109 & 0.004 & 27.069 & 0.592 \\ 
  & Stroop Color & 6 & -0.807 & 0.033 & 24.189 & 0.529 \\ 
  & Stroop Interference & 7 & -0.338 & 0.025 & 13.613 & 0.298 \\ 
   \hline
\end{tabular}
\caption{CCA: complete case analysis. ACE: ACE imputation. % CMI (conditional mean imputation), 
MCMI: multiple conditional mean imputation. cUHDRS: Composite Unified Huntington Disease Rating Scale. TMS: total motor score. SDMT: Symbol Digits Modality Test. TFC: total functional capacity. Scaled slope $= |\hat{\alpha} / {\rm SE}(\hat{\alpha})|$, a measure of how quickly the outcome changes over time until clinical diagnosis. Rank: ordering of scaled slopes estimated by the same method. Prop. to max.: proportion of scaled slope to the largest scaled slope estimated by the same method.}
\label{table:rank_slopes}
\end{table}

\subsection{Calculating sample size for a clinical trial}
Next, we calculate the sample size required for a hypothetical clinical trial of Huntington disease. Then, we compare the sample size estimates based on ACE imputation, conditional mean imputation, and complete case analysis. We present the sample size required to detect treatment effects in a clinical trial with SDMT score as the primary outcome. For simplicity, we assume a balanced design, such that both the treatment and placebo groups are of the same size. To investigate the treatment effect, we compare the trajectories of SDMT scores between the placebo (subscript $p$) and treatment (subscript $t$) groups, respectively, using the following linear models:
\bse
Y_p&=&\beta_p + \alpha_p (s - X) \\
Y_t&=&\beta_t + \alpha_t (s - X),
\ese
where $\beta$ denotes the intercept and $\alpha$ denotes the slope on time to clinical diagnosis, $s - X$. With these models, we aim to make inference about the quantity $\delta = \alpha_t - \alpha_p$. 

Specifically, to investigate whether the treatment under study slowed cognitive impairment (as measured by SDMT scores), we define the null and alternative hypotheses $H_0: \delta=0$ and $H_A:\delta > 0$, respectively. To calculate the required sample sizes to test these hypotheses with desired power, we begin with the Z-statistic for $\delta$, $\hat{\delta} \sqrt{ n_g \{ {\rm Var} (\delta) \}^{-1}}$. % \citep{guan2019sample}. 
From this statistic, the per-group sample size given type I error rate $\kappa$ and power $1 - \Gamma$ is
\bse
\left[ \frac{\sqrt{{\rm Var}(\delta)}\{\Phi^{-1}(1 - \Gamma)-\Phi^{-1}(\kappa)\}}{-d} \right]^2,
\ese
where $\Phi^{-1}(\cdot)$ denotes the inverse cumulative distribution function for the distribution $\Normal(0,1)$ and $d$ is the assumed true value of $\delta$. %$\Normal(0,1)$. 
For simplicity, we assume $\sqrt{{\rm Var}(\delta)}=1$.

Suppose that we are specifically interested in the sample size required to detect %$\alpha_t - \alpha_p = \delta > 0$ when there is 
a 10\% treatment effect with 80\% power, such that $\alpha_t = (1 - 0.1)\alpha_p$, $\kappa=0.05$, and $1 - \Gamma=0.8$. Borrowing the model estimates for $\alpha$ from Section~\ref{sec:ranking_outcomes} as the slopes on $s - X$ for the placebo group (i.e., using these estimates as $\alpha_p$) we can calculate this sample size based on each approach. Based on $\hat{\alpha_p}=-0.757$ from ACE imputation, we find that a per-group sample size of $1079$ would be needed for our desired test. In contrast, basing the calculations on the complete case analysis led to a smaller sample size ($\hat{\alpha_p}= -1.258$, 391 required) and MCMI led to a larger required sample size ($\hat{\alpha_p}= -0.739$, 1134 required). 

Given the theoretical and empirical evidence supporting ACE imputation, we believe $1079$ to be the most reliable per-group sample size for the proposed test. Hence, using this sample size as a reference, the sample size from complete case analysis would drastically under-power our ability to detect a 10\% treatment effect. Although the sample size from MCMI is much closer to that from ACE imputation, recruiting 55 more subjects per group would be both costly and potentially infeasible given the low prevalence of Huntington disease, at 12 per 100,000 people \citep{wexler2016george}. The corresponding power curves for all approaches for 10\%, 15\%, and 20\% treatment effects are given in Figure~\ref{fig:power_curves} in Appendix~\ref{sec:app_tables}. 

\section{Discussion}
\label{sec:discussion}

Because Huntington disease causes such multifaceted impairment, clinicians must select from many possible outcomes when designing clinical trials. To inform this selection, analysts have sought to rank possible outcomes by how rapidly they decline over time in gene mutation carriers. This ranking can only be done when we have accurate disease progression models, which rely on knowing the time at which subjects are clinically diagnosed with Huntington disease. When this time of clinical diagnosis is censored --- as it often is in observational studies of Huntington disease --- we can impute it, but then our outcome model estimates are sensitive to whether we correctly specified our imputation model; when we misspecify this imputation model, bias and inefficiency can result.

In this paper, we presented ACE imputation, which accurately estimates a longitudinal model given an imputed covariate, even when we use a misspecified imputation model. %In the first stage of our two-stage method, we leverage conditional mean imputation \citep{Atem2019, correctingCMI} to impute the censored covariate. In the second stage, we treat these imputed values as error-prone versions of the true covariates \citep{Carrolletal2006, haber2020bias} and implement an estimating equation which estimates the parameters of interest while actively correcting for this imputation error. 
We proved that this estimator can be calculated without positing distributions for the random effects $\bb$ or the imputation error $U$. Moreover, this estimator achieves consistency and asymptotic normality without prior knowledge about these nuisance distributions. Not having to estimate or postulate distributions for $\bb$ or $U$ is unlike typical semiparametric methods, which require positing nuisance distributions and which maintain consistency but suffer from inefficiency when the posited distributions are incorrect.

Although ACE imputation produces an estimator which is identifiable, consistent, and asymptotically normal, this novel method is not without limits. 
% \tgcomment{ADD COMMENTS HERE}
As we show in Appendix~\ref{sec:limitation}, the proposed estimating equation may fail to converge when a column of the fixed effects design matrix belongs to the column space of the random effects design matrix for all subjects. For example, this ``column space issue'' can arise when we include both a fixed intercept and a random intercept in our model. While more rigorous details are shown in Appendix~\ref{sec:limitation}, we can heuristically understand this issue as follows: when, for example, we include both fixed and random intercepts, the fixed intercept is the mean of the random intercept distribution. Therefore, there is a conflict when we simultaneously attempt to estimate the fixed intercept and treat the distribution of the random intercept (of which the fixed intercept is the mean) as a nuisance parameter. Given this understanding, perhaps this column space issue could be overcome by requiring the nuisance distribution of the random effects to have mean $\bzero$ so that we can untangle the fixed effects of interest from the nuisance distribution. Future investigations into solving this issue are required.

\section*{Supplementary Material} The supplementary material contains all technical details of our theorems and propositions. An open-source R package ACEimpute and code to replicate the simulation study is found at https://github.com/Tanya-Garcia-Lab/ACEimpute.

\bibliographystyle{apalike}

\bibliography{ms.bib}

\appendix

\bigskip
\bigskip
\bigskip
\begin{center}
    {\LARGE\bf Appendices}
\end{center}
\medskip

\section{Efficient Score Vectors to Estimate $\btheta$ in the Presence of Nuisance Parameters}
\label{sec:original_score_vectors}

%To estimate $\btheta$  under the most flexible modeling assumptions, we will not make any distributional assumptions on the random intercept, $\bb_i$, and imputation error, $U_i$.   Instead, treating the random intercept and imputation error  as so-called latent variables (i.e., a variable not observed), our longitudinal mixed effects model belongs in the class of generalized linear latent variable models.  For such a class of models, semiparametric methods exist \citep{garcia2016optimal} for which model parameters are estimated using an intermediate quantity that plays a similar role as the classical sufficient and complete statistic. This method results in parameter estimators that are consistent, efficient, and robust to misspecification. 

To apply the semiparametric framework from \citet{garcia2016optimal}, observe that the density for the $i$th subject in our model is:
\bse
f_{\bY_i,\tilD_i,{\rm mis}}(\bY_i,\tilD_i;\btheta)= \int f_{\bY_i|\tilD_i,\bb_i}(\bY_i|\tilD_i,\bb_i;\btheta)f_{\tilD_i,\bb_i,{\rm mis}}(\tilD_i,\bb_i)d\mu(\bb_i)
\ese
if $\Delta_i=1$ and \bse
f_{\bY_i,\widehat{\tilD}_i,{\rm mis}}(\bY_i,\widehat{\tilD}_i;\btheta)= \int f_{\bY_i|\widehat{\tilD}_i,\bb_i,U_i}(\bY_i|\widehat{\tilD}_i,\bb_i,U_i;\btheta)f_{\widehat{\tilD}_i,\bb_i,U_i,{\rm mis}}(\widehat{\tilD}_i,\bb_i,U_i)d\mu(\bb_{i}, U_i)
\ese
if $\Delta_i=0$. Here, the subscript $_{\rm{mis}}$ is used to indicate a term that is subject to misspecification due to misspecification of the joint density for $(\tilD_i, \bb_i)$ (or $(\tilD_i, \bb_i, U_i)$). Then, using similar techniques to those in \cite{garcia2016optimal}, we have that the efficient score vector for uncensored subjects  is:
\be
\label{eqn:orig_score_vec_uncens}
\bS_{\rm eff, mis}(\bY_i,\tilD_i;\btheta)=\bS_{\rm \btheta, mis}(\bY_i,\tilD_i;\btheta)-\E_{\rm{mis}}\{\bh(\tilD_i,\bb_i)|\bY_i,\tilD_i\}
\ee
where $\bS_{\rm \btheta, mis}(\bY_i,\tilD_i;\btheta)=\frac{\partial}{\partial \btheta}\log\{f_{\bY_i,\tilD_i,{\rm mis}}(\bY_i,\tilD_i;\btheta)\}$. Here, $\bh(\tilD_i,\bb_i)$ is a function with mean $\bzero$ and which satisfies
\bse
&&\E\{\bS_{\btheta, \rm{mis}}(\bY_i,\tilD_i;\btheta)|\tilD_i,\bb_i\}=\E[\E_{\rm{mis}}\{\bh(\tilD_i,\bb_i)|\bY_i,\tilD_i\}|\tilD_i,\bb_i].
\ese
Similarly, the efficient score vector for censored subjects is
\be
\label{eqn:orig_score_vec_cens}
\bS_{\rm eff, mis}^*(\bY_i,\widehat{\tilD}_i;\btheta)=\bS_{\rm \btheta, mis}^*(\bY_i,\widehat{\tilD}_i;\btheta)-\E_{\rm{mis}}\{\bh(\widehat{\tilD}_i,\bb_i,U_i)|\bY_i,\widehat{\tilD}_i\}
\ee
where $\bS_{\rm \btheta, mis}^*(\bY_i,\widehat{\tilD}_i;\btheta)=\frac{\partial}{\partial \btheta}\log\{f_{\bY_i,\widehat{\tilD}_i,{\rm mis}}(\bY_i,\widehat{\tilD}_i;\btheta)\}$ and  $\bh(\widehat{\tilD}_i,\bb_i,U_i)$ is a function with mean $\bzero$ and which satisfies
\bse
&&\E\{\bS_{\btheta, \rm{mis}}^*(\bY_i,\widehat{\tilD}_i;\btheta)|\widehat{\tilD}_i,\bb_i,U_i\}=\E[\E_{\rm{mis}}\{\bh(\widehat{\tilD}_i,\bb_i,U_i)|\bY_i,\widehat{\tilD}_i\}|\widehat{\tilD}_i,\bb_i,U_i].
\ese

Yet, as \cite{garcia2016optimal} point out, it is difficult to find functions  $\bh(\tilD_i,\bb_i)$ and $\bh(\widehat{\tilD}_i,\bb_i,U_i)$ that satisfy these conditions. In Section~\ref{sec:simplify_score_vector}, we simplify the efficient score vectors given in Equation~\eqref{eqn:orig_score_vec_uncens} and Equation~\eqref{eqn:orig_score_vec_cens}. We achieve this simplification by leveraging the properties of multivariate normal distributions and linear projection theory. In so doing, we circumvent the need for the functions $\bh(\tilD_i,\bb_i)$ and $\bh(\widehat{\tilD}_i,\bb_i,U_i)$ altogether and provide straightforward, closed forms of both efficient score vectors. Moreover, we show in Appendix~\ref{sec:calcs} that these efficient score vectors can be calculated \textbf{without} specifying joint distributions for $(\tilD, \bb_i)$ or $(\widehat{\tilD}, \bb_i, U_i)$; this lack of specification is impactful because if we did have to specify these joint distributions, we would run the risk of misspecifying them, which would then decrease the overall efficiency of $\ACEest$, our estimator for $\btheta$.

\section{Simplifying the Efficient Score Vectors}
\label{sec:simplify_score_vector}

\subsection{Transforming the Response Vector for Uncensored Subjects}
\label{sec:simplify_score_vector_uncens}
To simplify the efficient score vector for uncensored subjects, we will 1) transform $\bY_i$ for censored subjects, 2) establish five properties about the conditional distribution of $\bY_i$ given this transformation and $(\tilD_i,\bb_i)$, then 3) leverage these properties to simplify the complicated form of the efficient score vector given in Equation~\eqref{eqn:orig_score_vec_uncens} above. Specifically, we transform $\bY_i$ via $\bT_i=\sigma^{-2}\bZ^{bT}_i\bY_i$. In Proposition~\ref{prop:three_properties}, we outline the five key properties pertaining to the conditional distribution of $\bY_i$ given $(\bT_i,\tilD_i,\bb_i)$:

\begin{proposition}
\label{prop:three_properties}
Consider the transformed response variable denoted by $\bT_i=\sigma^{-2}\bZ^{bT}_i\bY_i$. If the random effects design matrix $\bZ^{b}_i$ is of full column rank (i.e., if ${\rm rank}(\bZ^{b}_i)=p_b$), then,
\begin{enumerate}
    \item[(A)] $\rm{E}(\bY_i|\bT_i,\tilD_i,\bb_i)=\bzeta(\tilD_i;\btheta)+\bP_{\bZ_i^b}\{\bY_i-\bzeta(\tilD_i;\btheta)\}$, where $\bP_{\bM} = \bM(\bM^T\bM)^{-1}\bM^T$ is the orthogonal projection operator for a matrix $\bM$.
    \item[(B)] Conditional on $(\bT_i,\tilD_i)$, $\bY_i$ and $\bb_i$ are independent.
    \item[(C)] $\rm{E}(\bY_i^T\bY_i|\bT_i,\tilD_i,\bb_i)=\sigma^2(m_i-p_b)+||\rm{E}(\bY_i|\bT_i,\tilD_i,\bb_i)||^2_2$, where $||\cdot||^2_2$ denotes the L2 norm.
    \item[(D)] For any vector-valued function $\bg(\bT_i,\tilD_i)$, if $\E\{\bg(\bT_i,\tilD_i)|\tilD_i,\bb_i\}=\bzero$, then it follows that $\bg(\bT_i,\tilD_i)=\bzero$.
    \item[(E)] Given a random variable $\bR$, $\bR$ is conditionally independent of $\bT_i=\sigma^{-2}\bZ^{bT}_i\bY_i$ given $(\bY_i, \bZ_i^b)$.
\end{enumerate}
\end{proposition}

To prove Proposition~\ref{prop:three_properties}, we first return to the outcome model of interest:
\bse
&&Y_{ij} =
\bbeta^T \bZ^a_{ij} + \alpha(s_{ij} - X_i) + \bb_i^T \bZ^b_{ij} + \epsilon_{ij} \\
&&\bb_i\sim f_{\bb}, \ \ \ \epsilon_{ij}\sim\Normal(0,\sigma^2).
\ese
From this model, we can see that conditional on $(\bZ^a_{ij}, s_{ij}, X_i, \bZ^b_{ij}, \bb_i)$, the response $\bY_{ij}$ follows a univariate normal distribution, $\Normal\{\bbeta^T \bZ^a_{ij} + \alpha(S_{ij} - X_i) + \bZ^b_{ij}\bb_i, \sigma^2\}$. Recall that we assume that responses from the $i$th subject are independent conditional on that subject's random effects $\bb_i$. Given this fact and the conditional distribution of $Y_{ij}$, we know that, conditional on $(\tilD_i, \bb_i)$, the $m$-dimensional response vector $\bY_{i}$ follows a multivariate normal distribution, $\Normal_{m_i}\{\bZ^a_i\bbeta  + \alpha(\bS_i - X_i\bone_{m_i}) + \bZ^b_i\bb_i, \sigma^2\bI_{m_i}\}$, where $\bone_{m_i}$ denotes an $m_i$-vector of ones and $\bI_{m_i}$ denotes the $m_i \times m_i$ identity matrix. For notational simplicity, let $\bzeta(\tilD_i;\btheta)=\bZ^a_i\bbeta  + \alpha(\bS_i - X_i\bone_{m_i})$. We consider this to be the fixed effects component of the linear predictor for $\bY_i$.

Next, we derive the joint conditional distribution of $(\bY_i, \bT_i)$ given $(\tilD_i,\bb_i)$. Given our definition of $\bT_i=\sigma^{-2}\bZ^{bT}_i\bY_i$, it follows that $(\bY_i^T, \bT_i^T)^T = (\bI_{m_i}, \sigma^{-2}\bZ_i^b)^T \bY_i$. From this, we know that, conditional on $(\tilD_i,\bb_i)$, $(\bY_i, \bT_i)\sim\Normal_{m_i}[(\bI_{m_i}, \sigma^{-2}\bZ_i^b)^T\{\bzeta(\tilD_i;\btheta) + \bZ_i^b\bb_i\}, \sigma^2(\bI_{m_i}, \sigma^{-2}\bZ_i^b)^T(\bI_{m_i}, \sigma^{-2}\bZ_i^b)]$. Specifically,
\be
\left(
\begin{matrix}
\bI_{m_i} \\
\sigma^{-2}\bZ_i^{bT}
\end{matrix}
\right)\{
\bzeta(\tilD_i;\btheta) + \bZ_i^b\bb_i\}
\nonumber&=& 
\left(
\begin{matrix}
\bzeta(\tilD_i;\btheta) + \bZ_i^b\bb_i \\
\sigma^{-2}\bZ_i^{bT}\{\bzeta(\tilD_i;\btheta) + \bZ_i^b\bb_i\}
\end{matrix}
\right) \label{eqn:YW-distr} \\
\sigma^2 
\left(
\begin{matrix}
\bI_{m_i} \\
\sigma^{-2}\bZ_i^{bT}
\end{matrix}
\right)
\left(
\begin{matrix}
\bI_{m_i} &
\sigma^{-2}\bZ_i^b
\end{matrix}
\right) &=&
\sigma^2
\left(
\begin{matrix}
\bI_{m_i} & \sigma^{-2}\bZ_i^b \\
\sigma^{-2}\bZ_i^{bT} & 
\sigma^{-4}\bZ_i^{bT}\bZ_i^b
\end{matrix}
\right).\nonumber
\ee
Next, we establish the conditional distribution of $\bY_i$ given $(\bT_i, \tilD_i,\bb_i)$. Using the previous result, we know that $\bY_i$ given $(\bT_i, \tilD_i,\bb_i)$ is also normally distributed with
\bse
\rm{E}(\bY_i|\bT_i,\tilD_i,\bb_i)&=&\bzeta(\tilD_i;\btheta)+\bZ_i^b\bb_i+\sigma^2\bZ_i^b(\bZ_i^{bT}\bZ_i^b)^{-1}[\bT_i - \sigma^{-2}\bZ_i^{bT}\{\bzeta(\tilD_i;\btheta) + \bZ_i^b\bb_i\}]\\
&=& \sigma^2\bZ_i^b(\bZ_i^{bT}\bZ_i^b)^{-1}\bT_i + (\bI_{m_i} - \bP_{\bZ_i^b})\{\bzeta(\tilD_i;\btheta) + \bZ_i^b\bb_i\},
\ese
where $\bP_{\bZ_i^b} = \bZ_i^b(\bZ_i^{bT}\bZ_i^b)^{-1}\bZ_i^{bT}$ is the orthogonal projection operator onto the column space of $\bZ_i^b$. Using this, we note that  $(\bI_{m_i} - \bP_{\bZ_i^b})\bZ_i^b\bb_i=0$ since $\bP_{\bZ_i^b}\bZ_i^b = \bZ_i^b$, given our assumption that$\bZ_i^b$ is of full column rank. Thus, we conclude that
\bse
\rm{E}(\bY_i|\bT_i,\tilD_i,\bb_i)&=&
\sigma^2\bZ_i^b(\bZ_i^{bT}\bZ_i^b)^{-1}\bT_i + (\bI_{m_i} - \bP_{\bZ_i^b})\bzeta(\tilD_i;\btheta)\\
&=&\bZ_i^b(\bZ_i^{bT}\bZ_i^b)^{-1}\bZ^{bT}_i\bY_i + (\bI_{m_i} - \bP_{\bZ_i^b})\bzeta(\tilD_i;\btheta)\\
&=&\bzeta(\tilD_i;\btheta)+\bP_{\bZ_i^b}\{\bY_i-\bzeta(\tilD_i;\btheta)\},
\ese
and so Proposition~\ref{prop:three_properties} (A) is established. Next, observe that
\bse
\rm{Cov}(\bY_i|\bT_i,\tilD_i,\bb_i)&=&\sigma^2\bI_{m_i} - \sigma^2\bZ_i^b(\bZ_i^{bT}\bZ_i^b)^{-1}\bZ_i^{bT} =\sigma^2(\bI_{m_i} - \bP_{\bZ_i^b}).
\ese
In summary,
\bse
\bY_i | (\bT_i,\tilD_i) \sim \Normal_{m_i}[\bzeta(\tilD_i;\btheta)+\bP_{\bZ_i^b}\{\bY_i-\bzeta(\tilD_i;\btheta)\}, \sigma^2(\bI_{m_i} - \bP_{\bZ_i^b})].
\ese
Observe that this distribution does not involve the random effects $\bb_i$. We therefore conclude --- because a multivariate normal distribution is completely specified by its mean and covariance --- that $\bY_i$ and $\bb_i$ are conditionally independent given $(\bT_i,\tilD_i)$. This confirms Proposition~\ref{prop:three_properties} (B). 

% \textbf{FACT:} For an orthogonal projection operator $\bP$ of rank $r$, $\bI_m - \bP$ is itself an orthogonal projection operator of rank $m-r$.

Next, we note that for a random vector $\bY$ with mean $\bmu$ and variance-covariance matrix $\bSigma$, $E(\bY^T\bY) = \bmu^T\bmu + tr(\bSigma)$. Given this, it follows from the conditional distribution of $\bY_i$ given $(\bT_i, \tilD_i,\bb_i)$ that 
\bse
\rm{E}(\bY_i^T\bY_i|\bT_i,\tilD_i,\bb_i)&=&\rm{tr}\{\sigma^2(\bI_{m_i}-\bP_{\bZ_i^b})\}+||\rm{E}(\bY_i|\bT_i,\tilD_i,\bb_i)||^2_2\\
&=&\sigma^2\rm{tr}(\bI_{m_i}-\bP_{\bZ_i^b})+||\rm{E}(\bY_i|\bT_i,\tilD_i,\bb_i)||^2_2\\
&=&\sigma^2(m_i-p_b)+||\rm{E}(\bY_i|\bT_i,\tilD_i,\bb_i)||^2_2,
\ese
% $\rm{E}(\bY_i^T\bY_i|\bT_i,\tilD_i,\bb_i)=\rm{tr}\{\sigma^2(\bI_{m_i}-\bP_{\bZ_i^b})\}+||\rm{E}(\bY_i|\bT_i,\tilD_i,\bb_i)||^2_2$. For simplicity, note that $\rm{tr}\{\sigma^2(\bI_{m_i}-\bP_{\bZ_i^b})\} = \sigma^2\rm{tr}(\bI_{m_i}-\bP_{\bZ_i^b}) = \sigma^2(m_i-p_b)$
since $\bI_m - \bP$ is itself an orthogonal projection operator of rank $m-r$ and the trace of an orthogonal projection operator is equal to its rank. This establishes Proposition~\ref{prop:three_properties} (C).

Proposition~\ref{prop:three_properties} (D) follows from the following calculation, given an arbitrary function $\bg(\bT_i,\tilD_i)$ such that $\E\{\bg(\bT_i,\tilD_i)|\bb_i,\tilD_i\}=\bzero$:
\bse
&&\bzero=\E\{\bg(\bT_i,\tilD_i)|\tilD_i,\bb_i\}\\
&=& \int \bg(\bT_i,\tilD_i)f(\bT_i|\tilD_i,\bb_i)d\mu(\bT_i)\\
&=& \int \bg(\bT_i,\tilD_i)
\exp\Big[-\frac{\sigma^4}{2}\{\bT_i-\sigma^{-2}\bZ_i^{bT}(\bzeta (\tilD_i;\btheta)+\bZ_i^b\bb_i)\}^T\times\\
&&(\bZ_i^{bT}\bZ_i^{b})^{-1}\{\bT_i-\sigma^{-2}\bZ_i^{bT}(\bzeta(\tilD_i;\btheta) +\bZ_i^b\bb_i)\}\Big]d\mu(\bT_i),
\ese
which follows from the conditional distribution of $(\bY_i, \bT_i)$ in Equation~\eqref{eqn:YW-distr}. Next, define
\bse
k(\bT_i,\tilD_i;\btheta)&=&\exp\Big[-\frac{\sigma^4}{2}\{\bT_i-\sigma^{-2}\bZ_i^{bT}(\bzeta (\tilD_i;\btheta)+\bZ_i^b\bb_i)\}^T\times\\
&&(\bZ_i^{bT}\bZ_i^{b})^{-1}\{\bT_i-\sigma^{-2}\bZ_i^{bT}(\bzeta(\tilD_i;\btheta) +\bZ_i^b\bb_i)\}\Big]d\mu(\bT_i),
\ese
which is positive for all $(\bT_i,\tilD_i,\btheta)$. Hence, the fact that
\bse
\bzero = \int k(\bT_i,\tilD_i;\btheta)
\bg(\bT_i,\tilD_i)d\mu(\bT_i)
\ese
for all $\btheta$ implies that $k(\bT_i,\tilD_i;\btheta)\bg(\bT_i,\tilD_i)=\bzero$. Yet, because $k(\bT_i,\tilD_i;\btheta)$ is positive for all $(\bT_i,\tilD_i,\btheta)$, this can only be true if $\bg(\bT_i,\tilD_i)=\bzero$. This establishes Proposition~\ref{prop:three_properties} (D). Proposition~\ref{prop:three_properties} (E) follows from Lemma~\ref{lem:cond_ind_of_func}:

\begin{Lem}
\label{lem:cond_ind_of_func}
Given random variables $\bR_1$, $\bR_2$, and $\bR_3$ and a vector-valued function $\bg(\cdot)$, $\bR_1$ is independent of $\bg(\bR_2, \bR_3)$ conditional on $(\bR_2, \bR_3)$.
\end{Lem}
To prove Lemma~\ref{lem:cond_ind_of_func}, observe:
\bse
&&P\{\bR_1 \leq \br_1, \bg(\bR_2,\bR_3) \leq \bk | \bR_2 = \br_2, \bR_3 = \br_3\} \\
&=&P(\bR_1 \leq \br_1 | \bR_2 = \br_2, \bR_3 = \br_3)I\{\bg(\br_2,\br_3) \leq \bk\} \\
&=&P(\bR_1 \leq \br_1 | \bR_2 = \br_2, \bR_3 = \br_3)P\{\bg(\bR_2,\bR_3) \leq \bk | \bR_2 = \br_2, \bR_3 = \br_3\}.
\ese

Proposition~\ref{prop:three_properties} (E) follows directly from Lemma~\ref{lem:cond_ind_of_func}, since $\bT_i = \sigma^{-2}\bZ_i^{bT}\bY_i$ is a function of $(\bY_i, \bZ_i^b)$. Hence, Proposition~\ref{prop:three_properties} is proven in full. 

\subsection{Transforming the Response Vector for Censored Subjects}
\label{sec:simplify_score_vector_cens}

Similarly, we will transform the response vector for censored subjects, establish five properties about the conditional distribution of the response given this transformation, then leverage these properties to simplify the complicated form of the efficient score vector given in Equation~\eqref{eqn:orig_score_vec_cens}. These five properties are outlined in Proposition~\ref{prop:three_properties_cens}:
\begin{proposition}
\label{prop:three_properties_cens}
Consider the transformed response variable $\bT_i^{\rm aug}=\sigma^{-2}(\bone_{m_i},\bZ_i^b)^T\bY_i$. If the matrix $(\bone_{m_i},\bZ_i^b)$ is of full column rank (i.e., if ${\rm rank}\{(\bone_{m_i},\bZ_i^b)\}=p_b+1$), then,
\begin{enumerate}
    \item[$\rm{P1}^*$] $\rm{E}(\bY_i|\bT_i^{\rm aug},\widehat{\tilD}_i,\bb_i,U_i)=\bzeta(\widehat{\tilD}_i;\btheta)+\bP_{(\bone_{m_i},\bZ_i^b)}\{\bY_i-\bzeta(\widehat{\tilD}_i;\btheta)\}$, where the matrix $\bP_{\bM} = \bM(\bM^T\bM)^{-1}\bM^T$ is the orthogonal projection operator for a matrix $\bM$.
    \item[$\rm{P2}^*$] Conditional on $(\bT_i^{\rm aug},\widehat{\tilD}_i)$, $\bY_i$ and $(\bb_i,U_i)$ are independent.
    \item[${\rm P3}^*$] ${\rm E}(\bY_i^T\bY_i|\bT_i^{\rm aug},\widehat{\tilD}_i,\bb_i,U_i)=\sigma^2(m_i-p_b)+||\rm{E}(\bY_i|\bT_i^{\rm aug},\widehat{\tilD}_i,\bb_i,U_i)||^2_2$, where $||\cdot||^2_2$ denotes the L2 norm.
    \item[${\rm P4}^*$] Given any function $\bg(\bT_i^{\rm aug},\widehat{\tilD}_i)$, if $\E\{\bg(\bT_i^{\rm aug},\widehat{\tilD}_i)|\widehat{\tilD}_i,\bb_i,U_i\}=\bzero$, then it follows that $\bg(\bT_i^{\rm aug},\widehat{\tilD}_i)=\bzero$.
    \item[${\rm P5}^*$] Given any random variable $\bR$, $\bR$ is conditionally independent of $\bT_i^{\rm aug}$ given $(\bY_i, \bZ_i^b)$.
\end{enumerate}
\end{proposition}

We begin the proof of Proposition~\ref{prop:three_properties_cens} but omit the rest because it follows analogously from the proof of Proposition~\ref{prop:three_properties}. Recall that for censored subjects, the outcome model is equivalent to
\bse
&&Y_{ij} =
\bbeta^T \bZ^a_{ij} + \alpha(s_{ij} - \widehat{X}_i - U_i) + \bb_i^T \bZ^b_{ij} + \epsilon_{ij}\\
&&\bb_i\sim f_{\bb}, \ \ \ U_i\sim f_U, \ \ \  \epsilon_{ij}\sim\Normal(0,\sigma^2)
\ese
since the imputed conditional mean $\widehat{X}_i$ is related to the censored $X_i$ through a Berkson error model, $X_i = \widehat{X}_i + U_i$. Then, conditional on the variables $(\widehat{\tilD}_i, \bb_i,U_i)$, the $m_i$-dimensional response vector $\bY_{i}$ follows a multivariate normal distribution $\Normal_{m_i}\{\bZ^a_i\bbeta  + \alpha(\bS_i - \widehat{X}_i\bone_{m_i} - U_i\bone_{m_i}) + \bZ^b_i\bb_i, \sigma^2\bI_{m_i}\}$. To simplify the mean of this multivariate normal distribution, we reuse the notation $\bzeta(\widehat{\tilD}_i;\btheta)=\bZ^a_i\bbeta+\alpha(\bs_i - \widehat{X}_i\bone_{m_i})$. With this notation, the conditional mean of $\bY_i$ given $(\widehat{\tilD}_i, \bb_i,U_i)$ is now $\bzeta(\widehat{\tilD}_i;\btheta)+(\bone_{m_i},\bZ^b_i)(-\alpha U_i, \bb_i^T)^T$.
% do we need \bA?? There's already a lot of notation!!
% \bse
% \bA =
% \left(
% \begin{matrix}
% -\alpha & \bzero_q^T \\
% \bzero_q & I_q
% \end{matrix}
% \right)
% \ese
% such that $\bA(\bb_i,U_i) = (-\alpha U_i, \bb_i^T)^T$. 
Similar to Section~\ref{sec:simplify_score_vector_uncens}, we propose transforming the response variables for censored subjects via $\bT_i^{\rm aug}=\sigma^{-2}(\bone_{m_i},\bZ^b_i)^T\bY_i$. This is the same transformation used for the response vector for uncensored subjects, except that $\bZ_i^b$ has been augmented by a vector of ones (hence the ``aug'' superscript). From here, the proof of Proposition~\ref{prop:three_properties_cens} follows in paralell from that of Proposition~\ref{prop:three_properties}, with $\tilD_i$ replaced by $\widehat{\tilD}_i$, $\bZ_i^b$ replaced by $(\bone_{m_i},\bZ^b_i)$, $\bb_i$ replaced by $(\bb_i, U_i)$, and $\bT_i$ replaced by $\bT^{\rm aug}_i$.

\subsection{Finding a Closed Form for the Efficient Score Vectors}
\label{sec:simplify}
To simplify the efficient score vector for uncensored subjects (Equation~\eqref{eqn:orig_score_vec_uncens}), first recall that the original form of this efficient score vector requires that we find a function $\bh(\tilD_i,\bb_i)$ that satisfies
\bse
\E\{\bS_{\btheta, \rm{mis}}(\bY_i,\tilD_i;\btheta)|\tilD_i,\bb_{i}\}=\E[\E_{\rm{mis}}\{\bh(\tilD_i,\bb_{i})|\bY_i,\tilD_i\}|\tilD_i,\bb_{i}],
\ese
where $\bS_{\btheta,\rm{mis}}(\bY_i,\tilD_i)=\frac{\partial}{\partial \btheta}\log f_{\bY_i,\tilD_i,\rm{mis}}(\bY_i,\tilD_i;\btheta)$. Observe that
\bse
&&\E\{\bS_{\btheta, \rm{mis}}(\bY_i,\tilD_i;\btheta)|\tilD_i,\bb_i\}
=E[\E\{\bS_{\btheta, \rm{mis}}(\bY_i,\tilD_i;\btheta)|\bT_i,\tilD_i,\bb_i\}|\tilD_i,\bb_i]
\ese
by the law of total expectation. Furthermore,
\bse
&&\E[\E_{\rm{mis}}\{\bh(\tilD_i,\bb_i)|\bY_i,\tilD_i\}|\tilD_i,\bb_i]=\E[\E_{\rm{mis}}\{\bh(\tilD_i,\bb_i)|\bY_i,\bT_i,\tilD_i\}|\tilD_i,\bb_i]
\ese
by Proposition~\ref{prop:three_properties} (E). Therefore, our condition for $\bh(\tilD_i,\bb_{i})$ becomes
\bse
&&\E[\E\{\bS_{\btheta, \rm{mis}}(\bY_i,\tilD_i;\btheta)|\bT_i,\tilD_i,\bb_i\}|\tilD_i,\bb_i]=
\E[\E_{\rm{mis}}\{\bh(\tilD_i,\bb_i)|\bY_i,\bT_i,\tilD_i\}|\tilD_i,\bb_i].
\ese
We then leverage Proposition~\ref{prop:three_properties} (B) to arrive at the following two results: 
\bse
\E\{\bS_{\btheta, \rm{mis}}(\bY_i,\tilD_i;\btheta)|\bT_i,\tilD_i,\bb_i\}&=&\E\{\bS_{\btheta, \rm{mis}}(\bY_i,\tilD_i;\btheta)|\bT_i,\tilD_i\}\\
\E_{\rm{mis}}\{\bh(\tilD_i,\bb_i)|\bY_i,\bT_i,\tilD_i\}&=&
\E_{\rm{mis}}\{\bh(\tilD_i,\bb_i)|\bT_i,\tilD_i\}.
\ese
Combining these results, the efficient score must therefore satisfy
\bse
&&\E[\E\{\bS_{\btheta, \rm{mis}}(\bY_i,\tilD_i;\btheta)|\bT_i,\tilD_i\}|\tilD_i,\bb_i]=\E[\E_{\rm{mis}}\{\bh(\tilD_i,\bb_i)|\bT_i,\tilD_i\}|\tilD_i,\bb_i].
\ese
Together with Proposition~\ref{prop:three_properties} (D), this implies that 
\bse
\E\{\bS_{\btheta, \rm{mis}}(\bY_i,\tilD_i;\btheta)|\bT_i,\tilD_i\}=\E_{\rm{mis}}\{\bh(\tilD_i,\bb_i)|\bT_i,\tilD_i\},
\ese
which further implies that
\bse
&&\E_{\rm{mis}}\{\bh(\tilD_i,\bb_i)|\bY_i,\tilD_i\}=\E_{\rm{mis}}\{\bh(\tilD_i,\bb_i)|\bY_i,\bT_i,\tilD_i\}\\
&=&\E_{\rm{mis}}\{\bh(\tilD_i,\bb_i)|\bT_i,\tilD_i\}=\E\{\bS_{\btheta, \rm{mis}}(\bY_i,\tilD_i;\btheta)|\bT_i,\tilD_i\}
\ese
by Proposition~\ref{prop:three_properties} (E). The efficient score vector can therefore be written as
\be
\bS_{\rm eff,mis}(\bY_i,\tilD_i;\btheta)&=&
\bS_{\btheta, \rm{mis}}(\bY_i,\tilD_i;\btheta)-\E_{\rm{mis}}\{\bh(\tilD_i,\bb_i)|\bY_i,\tilD_i\}\nonumber\\
&=&\bS_{\btheta, \rm{mis}}(\bY_i,\tilD_i;\btheta)-\E\{\bS_{\btheta, \rm{mis}}(\bY_i,\tilD_i;\btheta)|\bT_i,\tilD_i\}\label{eqn:closed_form}.
\ee
Importantly, this is a closed form solution. We can simplify the first term in Equation~\eqref{eqn:closed_form}: % proof for this in Kyle's notes, 'RE Imp: relating score vectors'
\bse
\bS_{\btheta, \rm{mis}}(\bY_i,\tilD_i;\btheta)&=&\E_{\rm{mis}}\{\bS_{\btheta}(\bY_i,\tilD_i,\bb_i;\btheta)|\bY_i,\tilD_i\},
\ese
where the ``model'' score vector $\bS_{\btheta}(\bY_i,\tilD_i,\bb_i;\btheta)=\frac{\partial}{\partial\btheta} \log f_{\bY_i|\tilD_i,\bb_i}(\bY_i|\tilD_i,\bb_i)$, which is not subject to misspecification. We can also simplify the second term in Equation~\eqref{eqn:closed_form}:
\bse
\E\{\bS_{\btheta, \rm{mis}}(\bY_i,\tilD_i;\btheta)|\bT_i,\tilD_i\}&=&\E[\E_{\rm{mis}}\{\bS_{\btheta}(\bY_i,\tilD_i,\bb_i;\btheta)|\bY_i,\tilD_i\}|\bT_i,\tilD_i]\\
%&=&\int E_{\rm{mis}}\{\bS_{\btheta}(\bY,\tilD_i,\bb)|\bY,\tilD_i\} f(\bY|\bT,\tilD_i) d\mu(\by)\\
% &=&\int \left\{\int \bS_{\btheta}(\bY_i,\tilD_i,\bb_i;\btheta)f_{\rm{mis}}(\bb_i|\bY_i,\tilD_i)d\mu(\bb_i)\right\} f(\bY_i|\bT_i,\tilD_i) d\mu(\bY_i)\\
%&=&\int \left\{\int \bS_{\btheta}(\bY_i,\tilD_i,\bb_i;\btheta)f(\bb_i|\bY_i,\bT_i,\tilD_i)d\mu(\bb_i)\right\} f(\bY_i|\bT_i,\tilD_i) d\mu(\by)\\
&=&\int \left\{\int \bS_{\btheta}(\bY_i,\tilD_i,\bb_i;\btheta)f_{\rm{mis}}(\bb_i|\bT_i,\tilD_i)d\mu(\bb_i)\right\} \times \\
&& f(\bY_i|\bT_i,\tilD_i) d\mu(\by_i)\\
&=&\int \bS_{\btheta}(\bY_i,\tilD_i,\bb_i;\btheta)f_{\rm{mis}}(\bY_i,\bb_i|\bT_i,\tilD_i)d\mu(\bb_i)d\mu(\by_i)\\
&=&\E_{\rm{mis}}\{\bS_{\btheta}(\bY_i,\tilD_i,\bb_i;\btheta)|\bT_i,\tilD_i\}.
\ese
This leads to the following
\bse
\bS_{\rm eff, mis}(\bY_i,\tilD_i;\btheta)&=&\E_{\rm{mis}}\{\bS_{\btheta}(\bY_i,\tilD_i,\bb_i;\btheta)|\bY_i,\tilD_i\}-\\
&&\E_{\rm{mis}}\{\bS_{\btheta}(\bY_i,\tilD_i,\bb_i;\btheta)|\bT_i,\tilD_i\}.
\ese
% Therefore, we will solve for $\btheta$ using the following estimating equation. % which is essentially taking the score function and ``subtracting away'' the nuisance effect.
% \bse
% \sum_{i=1}^n \bS_{\rm eff,mis}(\bY_i,\tilD_i;\btheta)=\bzero
% \ese

Mirroring the work above, it can be shown that this efficient score vector for censored subjects is equivalent to %$\bS_{\rm eff, mis}^*(\bY_i,\widehat{\tilD}_i;\btheta)=$
\bse
\bS_{\rm eff, mis}^*(\bY_i,\widehat{\tilD}_i;\btheta)&=&\E_{\rm{mis}}\{\bS_{\btheta}^*(\bY_i,\widehat{\tilD}_i,\bb_i,U_i;\btheta)|\bY_i,\widehat{\tilD}_i\}-\\
&&\E_{\rm{mis}}\{\bS_{\btheta}^*(\bY_i,\widehat{\tilD}_i,\bb_i,U_i;\btheta)|\bT_i^{\rm aug},\widehat{\tilD}_i\}.
\ese

\section{Calculating the Efficient Score Vectors}
\label{sec:calcs}
\subsection{Uncensored Subjects}
\label{sec:calcs_uncens}
Recall from Section~(\ref{sec:simplify_score_vector_uncens}) that, conditional on $(\tilD_i, \bb_i)$, the response vector corresponding to a single individual, $\bY_i$, follows a $\Normal_{m_i}(\bZ_i^a\bbeta  + \alpha(\bs_i - X_i\bone_{m_i}) + \bZ_i^b\bb_i, \sigma^2\bI_{m_i})$ distribution. Therefore, we know that the density for all observations from a single individual, $f_{\bY_i|\tilD_i,\bb_i}(\bY_i|\tilD_i,\bb_i;\btheta)=$:
\bse
&&(2\pi\sigma^2)^{-m_i/2}\exp\left[-\frac{||\bY_i-\{\bZ_i^a\bbeta  + \alpha(\bs_i - X\bone_{m_i}) + \bZ_i^b\bb_i\}||^2_2}{2\sigma^2}\right]
\ese
Therefore, the log-likelihood $\log\{f_{\bY_i|\tilD_i,\bb_i}(\cdot)\}=$
\be
\label{eqn:log_lik_uncens}
&&-\frac{m_i}{2}\log(2\pi\sigma^2)-\frac{||\bY_i-\{\bZ_i^a\bbeta  + \alpha(\bs_i - X_i\bone_{m_i}) + \bZ_i^b\bb_i\}||^2_2}{2\sigma^2}
\ee
\subsubsection{Efficient score vector for $\bbeta$}
\label{sec:beta_uncens}
From the log-likelihood in Equation~\eqref{eqn:log_lik_uncens}, we obtain
\bse
\bS_{\bbeta}(\bY_i,\tilD_i,\bb_i;\btheta)&\equiv&\frac{\partial}{\partial\bbeta}\log\{f_{\bY_i|\tilD_i,\bb_i}(\cdot)\}\\
&=&\frac{\bZ_i^{aT}[\bY_i-\{\bZ_i^a\bbeta  + \alpha(\bs_i - X_i\bone_{m_i}) + \bZ_i^b\bb_i\}]}{\sigma^2}\\
&=&\frac{\bZ_i^{aT}\bY_i-\bk_1(\tilD_i,\bb_i)}{\sigma^2},
\ese
where $\bk_1(\tilD_i,\bb_i)=\bZ_i^{aT}\{\bZ_i^a\bbeta  + \alpha(\bs_i - X_i\bone_{m_i}) + \bZ_i^b\bb_i\}$. By Proposition~\ref{prop:three_properties} parts (B) and (E), we have:
\bse
\E_{\rm{mis}}\{\bk_1(\tilD_i,\bb_i)|\bY_i,\tilD_i\}
=\E_{\rm{mis}}\{\bk_1(\tilD_i,\bb_i)|\bT_i,\tilD_i\}.
\ese
This implies that the efficient score vector for $\bbeta$, $\bS_{\rm{eff},\bbeta}(\bY_i,\tilD_i;\btheta)$, is equivalent to:
\bse
\bS_{\rm{eff},\bbeta}(\bY_i,\tilD_i;\btheta)
&=&\E_{\rm{mis}}\{\bS_{\bbeta}(\bY_i,\tilD_i,\bb_i;\btheta)|\bY_i,\tilD_i\}-\E_{\rm{mis}}\{\bS_{\bbeta}(\bY_i,\tilD_i,\bb_i;\btheta)|\bT_i,\tilD_i\}\\
&=&\sigma^{-2}\Big[\E_{\rm{mis}}\{\bZ_i^{aT}\bY_i-\bk_1(\tilD_i,\bb_i)|\bY_i,\tilD_i\}-\\
&&\E_{\rm{mis}}\{\bZ_i^{aT}\bY_i-\bk_1(\tilD_i,\bb_i)|\bT_i,\tilD_i\}\Big]\\
&=&\sigma^{-2}\{\E_{\rm{mis}}(\bZ_i^{aT}\bY_i|\bY_i,\tilD_i)-\E_{\rm{mis}}(\bZ_i^{aT}\bY_i|\bT_i,\tilD_i)\}\\
&=&\frac{\bZ_i^{aT}\{\bY_i-\E_{\rm{mis}}(\bY_i|\bT_i,\tilD_i)\}}{\sigma^2}.
\ese
Recall that the $_{\rm{mis}}$ subscript denotes expectations that are subject to possible misspecification. Here, because $\E_{\rm{mis}}(\bY_i|\bT_i,\tilD_i)$  depends only on the posited outcome model -- which we assume to be correct -- this expectation is not subject to misspecification; hence, we drop the $_{\rm{mis}}$ subscript as follows:
\bse
\bS_{\rm{eff},\bbeta}(\bY_i,\tilD_i;\btheta)
=\frac{\bZ_i^{aT}\{\bY_i-\E(\bY_i|\bT_i,\tilD_i)\}}{\sigma^2}
\ese
where, by Proposition~\ref{prop:three_properties}, $\rm{E}(\bY_i|\bT_i,\tilD_i,\bb_i)=\bzeta(\tilD_i;\btheta)+\bP_{\bZ_i^b}\{\bY_i-\bzeta(\tilD_i;\btheta)\}$ where $\bP_{\bM} = \bM(\bM^T\bM)^{-1}\bM^T$ for a matrix $\bM$, $\bT_i = \sigma^{-2}\bZ_i^{bT}\bY_i$, and $\bzeta(\tilD_i;\btheta)=\bZ_i^a\bbeta  + \alpha(\bS_i - X_i\bone_{m_i})$. With this in mind, we can further simplify $\bS_{\rm{eff},\bbeta}(\bY_i,\tilD_i;\btheta)$.
Observe:
\bse
\bS_{\rm{eff},\bbeta}(\bY_i,\tilD_i;\btheta)&=&\frac{\bZ_i^{aT}\{\bY_i-\bzeta(\tilD_i;\btheta)-\bP_{\bZ_i^b}\{\bY_i-\bzeta(\tilD_i;\btheta)\}}{\sigma^2} \\
&=&\frac{\bZ_i^{aT}(\bI_{m_i} - \bP_{\bZ_i^b})\{\bY_i-\bzeta(\tilD_i;\btheta)\}}{\sigma^2}.
\ese
\subsubsection{Efficient score for $\alpha$}
\label{sec:alpha_uncens}
The derivation of the efficient score for $\alpha$, which we denote by $\bS_{\rm{eff},\alpha}(\bY_i,\tilD_i;\btheta)$, is analogous to the derivation of $\bS_{\rm{eff},\bbeta}(\bY_i,\tilD_i;\btheta)$.
\bse
S_{\rm{eff},\alpha}(\bY_i,\tilD_i;\btheta)&=&\frac{(\bs_i-X_i\bone_{m_i})^T\{\bY_i-\E(\bY_i|\bT_i,\tilD_i)\}}{\sigma^2}\\
&=&\frac{(\bs_i-X_i\bone_{m_i})^T(\bI_{m_i}-\bP_{\bZ_i^b})\{\bY_i-\bzeta(\tilD_i;\btheta)\}}{\sigma^2}.
\ese
\subsubsection{Efficient score for $\sigma^2$}
\label{sec:sigma_uncens}
From the log-likelihood in Equation~\eqref{eqn:log_lik_uncens}, we obtain 
\bse
\bS_{\sigma^2}(\bY_i,\tilD_i,\bb_i;\btheta)&\equiv&\frac{\partial}{\partial\sigma^2}\log\{ f_{\bY_i|\tilD_i,\bb_i}(\cdot)\}\\
&=&\frac{||\bY_i-\{\bzeta(\tilD_i;\btheta) + \bZ_i^b\bb_i\}||^2_2}{2\sigma^4}-\frac{m_i}{2\sigma^2}\\
&=&\frac{\bY_i^T\bY_i}{2\sigma^4}-\frac{\bY_i^T\bzeta(\tilD_i;\btheta)}{\sigma^4}-\frac{\bY_i^T\bZ_i^b\bb_i}{\sigma^4}+\frac{||\bzeta(\tilD_i;\btheta)+\bZ_i^b\bb_i||^2_2}{2\sigma^4}-\frac{m_i}{2\sigma^2}\\
%&=&\frac{\bY^T\bY}{2\sigma^4}-\frac{\bY^T\bzeta}{\sigma^4}-\frac{\bw^T\bb}{\sigma^2}+\frac{||\bzeta+\bZ^b\bb||^2_2}{2\sigma^4}-\frac{m}{2\sigma^2}\\
&=&\frac{\bY_i^T\bY_i}{2\sigma^4}-\frac{\bY_i^T\bzeta(\tilD_i;\btheta)}{\sigma^4}-\frac{\bT_i^T\bb_i}{\sigma^2}+\frac{k_2(\tilD_i,\bb_i)}{2\sigma^2},
\ese
where $k_2(\tilD_i,\bb_i)=\sigma^{-2}||\bzeta(\tilD_i;\btheta_i)+\bZ_i^b\bb_i||^2_2-m_i$. By Proposition~\ref{prop:three_properties} parts (B) and (E), we have: 
\bse
\E_{\rm{mis}}\{k_2(\tilD_i,\bb_i)|\bY_i,\tilD_i\}
=\E_{\rm{mis}}\{k_2(\tilD_i,\bb_i)|\bT_i,\tilD_i\}.
\ese
Again by Proposition~\ref{prop:three_properties} parts (B) and (E), $\E_{\rm{mis}}\{\bT_i^T\bb_i|\bY_i,\tilD_i\}=$
\bse
&&\E_{\rm{mis}}(\bT_i^T\bb_i|\bY_i,\bT_i,\tilD_i)
=\bT_i^T\E_{\rm{mis}}(\bb_i|\bY_i,\bT_i,\tilD_i)\\
&=&\bT_i^T\E_{\rm{mis}}(\bb_i|\bT_i,\tilD_i)=\E_{\rm{mis}}(\bT_i^T\bb_i|\bT_i,\tilD_i).
\ese
It then follows that the efficient score for $\sigma^2$, which we denote by $S_{\rm{eff},\sigma^2}(\bY_i,\tilD_i;\btheta)$, is:
\bse
S_{\rm{eff},\sigma^2}(\bY_i,\tilD_i;\btheta)&=&\E_{\rm{mis}}\{\bS_{\sigma^2}(\bY_i,\tilD_i,\bb_i;\btheta)|\bY_i,\tilD_i\}-\\
&&\E_{\rm{mis}}\{\bS_{\sigma^2}(\bY_i,\tilD_i,\bb_i;\btheta)|\bT_i,\tilD_i\}\\
&=&\E_{\rm{mis}}\left\{\frac{\bY_i^T\bY_i}{2\sigma^4}-\frac{\bY_i^T\bzeta(\tilD_i;\btheta)}{\sigma^4}|\bY_i,\tilD_i\right\}-\\
&&\E_{\rm{mis}}\left\{\frac{\bY_i^T\bY_i}{2\sigma^4}-\frac{\bY_i^T\bzeta(\tilD_i;\btheta)}{\sigma^4}|\bT_i,\tilD_i\right\}\\
&=&\frac{1}{\sigma^4}\Big[\frac{1}{2}\left\{\bY_i^T\bY_i-\E_{\rm{mis}}(\bY_i^T\bY_i|\bT_i,\tilD_i)\right\}-\\
&&\bzeta(\tilD_i;\btheta)^T\{\bY_i-\E_{\rm{mis}}(\bY_i|\bT_i,\tilD_i)\}\Big],
\ese
where we can again disregard the subscript $_{\rm{mis}}$ in these conditional expectations because the conditional distribution of $\bY_i$ given $(\bT_i,\tilD_i)$ is not subject to misspecification. 

\subsubsection{Efficient score vector for $\btheta$}
Importantly, we have shown that each component of $\bS_{\rm{eff, mis}}(\bY_i,\tilD_i;\btheta)$ is \textbf{not} subject to mis-specification. To emphasize this point, we simply refer to the efficient score vector for $\btheta=(\bbeta, \alpha, \sigma^2)^T$ (corresponding to uncensored subjects) as $\bS_{\rm{eff}}(\bY_i,\tilD_i;\btheta)$. Combining these results, we find that
\be
\label{eqn:app_uncens_eff_score_vec}
\sigma^4\bS_{\rm{eff}}(\bY_i,\tilD_i;\btheta)=\left[
\begin{matrix}
\sigma^2\bZ_i^{aT}(\bI_{m_i} - \bP_{\bZ_i^b})\{\bY_i-\bzeta(\tilD_i;\btheta)\} \\
\sigma^2(\bs_i-X_i\bone_{m_i})^T(\bI_{m_i} - \bP_{\bZ_i^b})\{\bY_i-\bzeta(\tilD_i;\btheta)\} \\
\frac{1}{2}\{\bY_i^T\bY_i-\E(\bY_i^T\bY_i|\bT_i,\tilD_i)\}-\bzeta^T(\tilD_i;\btheta)\{\bY_i-\E(\bY_i|\bT_i,\tilD_i)\}
\end{matrix}
\right].
\ee
Recall from Proposition~\ref{prop:three_properties} parts (A) and (C) that
\bse
\rm{E}(\bY_i|\bT_i,\tilD_i,\bb_i)&=&\bzeta(\tilD_i;\btheta)+\bP_{\bZ_i^b}\{\bY_i-\bzeta(\tilD_i;\btheta)\}\\
\rm{E}(\bY_i^T\bY_i|\bT_i,\tilD_i)&=&\sigma^2(m_i-p_b)+||\rm{E}(\bY_i|\bT_i,\tilD_i)||^2_2,
\ese
where $\bP_{\bM} = \bM(\bM^T\bM)^{-1}\bM^T$ for a matrix $\bM$, $\bT_i = \sigma^{-2}\bZ_i^{bT}\bY_i$, and $\bzeta(\tilD_i;\btheta)=\bZ_i^a\bbeta  + \alpha(\bS_i - X_i\bone_{m_i})$.

\subsection{Censored Subjects}
\label{sec:calcs_cens}
Recall from Section~(\ref{sec:simplify_score_vector_cens}) that, conditional on $(\widehat{\tilD}_i, \bb_i,U_i)$, the response vector $\bY_i$, from one censored subject, follows a multivariate normal distribution $\Normal_{m_i}\{\bZ^a_i\bbeta  + \alpha(\bS_i - \widehat{X}_i\bone_{m_i} - U_i\bone_{m_i}) + \bZ^b_i\bb_i, \sigma^2\bI_{m_i}\}$. Therefore, we know that the density for all observations from a single individual, $f_{\bY_i|\widehat{\tilD}_i,\bb_i,U_i}(\bY_i|\widehat{\tilD}_i,\bb_i,U_i;\btheta)=$
\bse
&&(2\pi\sigma^2)^{-m_i/2}\exp\left[-\frac{||\bY_i-\{\bZ_i^a\bbeta  + \alpha(\bs_i - \widehat{X}_i\bone_{m_i}) + (\bone_{m_i},\bZ_i^b)(-\alpha U_i, \bb_i^T)^T\}||^2_2}{2\sigma^2}\right].
\ese
Therefore, the log-likelihood $\log\{f_{\bY_i|\widehat{\tilD_i},\bb_i,U_i}(\bY_i|\widehat{\tilD_i},\bb_i,U_i;\btheta)\}=$
\be
\label{eqn:log_lik_cens}
&&-\frac{m_i}{2}\log(2\pi\sigma^2)-\frac{||\bY_i-\{\bZ_i^a\bbeta  + \alpha(\bs_i - \widehat{X}_i\bone_{m_i}) + (\bone_{m_i},\bZ_i^b)(-\alpha U_i, \bb_i^T)^T\}||^2_2}{2\sigma^2}.
\ee
\subsubsection{Efficient score vector for $\bbeta$}
The derivation of the efficient score vector for $\bbeta$ corresponding to censored subjects mirrors that for uncensored subjects (Appendix \ref{sec:beta_uncens}). From this, we find that:
\bse
\bS_{\rm{eff},\bbeta}^*(\bY_i,\widehat{\tilD}_i;\btheta)
&=&\frac{\bZ_i^{aT}(\bI_{m_i} - \bP_{(\bone_{m_i},\bZ^b_i)})\{\bY_i-\bzeta(\widehat{\tilD}_i;\btheta)\}}{\sigma^2}.
\ese
\subsubsection{Efficient score for $\alpha$}
Next, we consider the efficient score for $\alpha$ corresponding to censored subjects.
From the log-likelihood in Equation~\eqref{eqn:log_lik_cens}, we find that: 
\bse
\bS_{\alpha}(\bY_i,\widehat{\tilD}_i,\bb_i,U_i;\btheta)&\equiv&\frac{\partial}{\partial\alpha}\log\{ f_{\bY_i|\widehat{\tilD}_i,\bb_i,U_i}(\bY_i|\widehat{\tilD}_i,\bb_i,U_i)\}\\
&=&\frac{\{\bS_i-(\widehat{X}_i+U_i)\bone_{m_i}\}^T[\bY_i-\{\bzeta(\widehat{\tilD}_i;\btheta) + (\bone_{m_i},\bZ_i^b)(-\alpha U_i, \bb_i^T)^T\}]}{\sigma^2}\\
&=&\frac{\{\bS_i-(\widehat{X}_i+U_i)\bone_{m_i}\}^T\bY_i-\bk_3(\widehat{\tilD}_i,\bb_i,U_i)}{\sigma^2},
\ese
where $\bk_3(\widehat{\tilD}_i,\bb_i,U_i)=\{\bS_i-(\widehat{X}_i+U_i)\bone_{m_i}\}^T\{\bzeta(\widehat{\tilD}_i;\btheta) + (\bone_{m_i},\bZ_i^b)(-\alpha U_i, \bb_i^T)^T\}$. Then, by again leveraging Proposition~\ref{prop:three_properties_cens} parts (B) and (E), we have:
\bse
\E_{\rm{mis}}\{\bk_3(\widehat{\tilD}_i,\bb_i,U_i)|\bY_i,\widehat{\tilD}_i\}
=\E_{\rm{mis}}\{\bk_3(\widehat{\tilD}_i,\bb_i,U_i)|\bT_i^{\rm aug},\widehat{\tilD}_i\}.
\ese
Next, we consider the term $\sigma^{-2}U_i\bone_{m_i}^T\bY_i$. Since we defined $\bT_i^{\rm aug}= \sigma^{-2}(\bone_{m_i},\bZ^b_i)^T\bY_i$, it follows that $\sigma^{-2}U_i\bone_{m_i}^T\bY_i=U_iT_{i1}^{\rm aug}$, where $T^{\rm aug}_{i1}$ denotes the first element of $\bT_i^{\rm aug}$. Then, it follows from Proposition~\ref{prop:three_properties_cens} parts (B) and (E) that $E(U_iT_{i1}^{\rm aug}|\bY_i,\widehat{\tilD}_i)=$
\bse
&&E(U_iT_{i1}^{\rm aug}|\bY_i,\bT_i^{\rm aug},\widehat{\tilD}_i)=T_{i1}^{\rm aug}E(U_i|\bY_i,\bT_i^{\rm aug},\widehat{\tilD}_i)\\
&=&T_{i1}^{\rm aug}E(U_i|\bT_i^{\rm aug},\widehat{\tilD}_i)=E(U_iT_{i1}^{\rm aug}|\bT_i^{\rm aug},\widehat{\tilD}_i).
\ese
Therefore, the conditional expectations of $\bk_3(\widehat{\tilD}_i,\bb_i,U_i)$ and $U_iT_{i1}^{\rm aug}$ drop out when calculating the efficient score for $\alpha$. As a result, we find that
\bse
S_{\rm{eff},\alpha}^*(\bY_i,\widehat{\tilD}_i;\btheta)=\frac{(\bs_i-\widehat{X_i}\bone_{m_i})^T(\bI_{m_i}-\bP_{(\bone_{m_i},\bZ^b_i)})\{\bY_i-\bzeta(\widehat{\tilD}_i;\btheta)\}}{\sigma^2}.
\ese
\subsubsection{Efficient score for $\sigma^2$}
The derivation of the efficient score for $\sigma^2$ corresponding to censored subjects mirrors the derivation shown in Appendix~(\ref{sec:sigma_uncens}). Hence, we find that:
\bse
S_{\rm{eff},\sigma^2}^*(\bY_i,\widehat{\tilD}_i;\btheta)
&=&\frac{1}{\sigma^4}\Big[\frac{1}{2}\left\{\bY_i^T\bY_i-\E(\bY_i^T\bY_i|\bT_i^{\rm aug},\widehat{\tilD}_i)\right\}-\\
&&\bzeta^T(\widehat{\tilD}_i;\btheta)\{\bY_i-\E(\bY_i|\bT_i^{\rm aug},\widehat{\tilD}_i)\}\Big].
\ese

\subsubsection{Efficient score vector for $\btheta$}
Combining these results, we find that
\be
\label{eqn:app_cens_eff_score_vec}
&&\sigma^4\bS_{\rm{eff}}^*(\bY_i,\widehat{\tilD}_i;\btheta)\nonumber\\
&=&\left[
\begin{matrix}
\sigma^2\bZ_i^{aT}(\bI_{m_i} - \bP_{(\bone_{m_i},\bZ^b_i)})\{\bY_i-\bzeta(\widehat{\tilD}_i;\btheta)\} \\
\sigma^2(\bs_i-\widehat{X}_i\bone_{m_i})^T(\bI_{m_i} - \bP_{(\bone_{m_i},\bZ^b_i)})\{\bY_i-\bzeta(\widehat{\tilD}_i;\btheta)\} \\
\frac{1}{2}\{\bY_i^T\bY_i-\E(\bY_i^T\bY_i|\bT_i^{\rm aug},\widehat{\tilD}_i)\}-\bzeta^T(\widehat{\tilD}_i;\btheta)\{\bY_i-\E(\bY_i|\bT_i^{\rm aug},\widehat{\tilD}_i)\}
\end{matrix}
\right].
\ee
Recall from Proposition~\ref{prop:three_properties_cens} parts (A) and (C) that
\bse
\rm{E}(\bY_i|\bT_i^{\rm aug},\widehat{\tilD}_i,\bb_i,U_i)&=&\bzeta(\widehat{\tilD}_i;\btheta)+\bP_{(\bone_{m_i},\bZ_i^b)}\{\bY_i-\bzeta(\widehat{\tilD}_i;\btheta)\}\\
E(\bY_i^T\bY_i|\bT_i^{\rm aug},\widehat{\tilD}_i)&=&\sigma^2(m_i-p_b)+||\rm{E}(\bY_i|\bT_i^{\rm aug},\widehat{\tilD}_i)||^2_2,
\ese
where $\bP_{\bM} = \bM(\bM^T\bM)^{-1}\bM^T$ for a matrix $\bM$, $\bT_i^{\rm aug} = \sigma^{-2}(\bone_{m_i}, \bZ^b)^T\bY_i$, and $\bzeta(\widehat{\tilD}_i;\btheta)=\bZ_i^a\bbeta  + \alpha(\bS_i - \widehat{X}_i\bone_{m_i})$.
\subsection{The Full Estimating Equation}
We have now derived the efficient score vectors for $\btheta$ corresponding to both uncensored and censored subjects. Combining these vectors gives us the total estimating equation:
\be
\label{eqn:app_full_estimating_equation}
\sum_{i=1}^n \left\{\Delta_i\bS_{\rm{eff}}(\bY_i,\tilD_i;\btheta) + (1 - \Delta_i)\bS_{\rm{eff}}^*(\bY_i,\widehat{\tilD}_i;\btheta)\right\}=\bzero,
\ee
where the censoring indicator $\Delta_i$ is $1$ if subject $i$ is not censored and $0$ otherwise. The efficient score vectors $\bS_{\rm{eff}}(\bY_i,\tilD_i;\btheta)$ and $\bS_{\rm{eff}}^*(\bY_i,\widehat{\tilD}_i;\btheta)$ are given in Equation~\eqref{eqn:app_uncens_eff_score_vec} and Equation~\eqref{eqn:app_cens_eff_score_vec}, respectively.

\section{Proof of Theorem~\ref{theorem:ident}}
\label{sec:ident_proof}

We now prove that $\ACEest$ is identifiable, i.e., that Theorem~\ref{theorem:ident} in the main text is true. 
We begin by considering the first $p + 1$ elements of this estimating equation (i.e., the elements corresponding to $(\bbeta, \alpha)$) in Equation~\eqref{eqn:total_estimating_equation}:
\bse
\bzero=&&\sum_{i=1}^n\Big[\Delta_i \sigma^2(\bZ^a_i,\bs_i-X_i\bone_{m_i})^T(\bI_m - \bP_{\bZ^b_i})\{\bY_i-(\bZ^a_i,\bs_i-X_i\bone_{m_i})(\bbeta^T, \alpha)^T\}+\\
&&(1 - \Delta_i)\sigma^2(\bZ^a_i,\bs_i-\widehat{X}_i\bone_{m_i})^T(\bI_m - \bP_{(\bone_{m_i},\bZ^b_i)})\{\bY_i-(\bZ^a_i,\bs_i-\widehat{X}_i\bone_{m_i})(\bbeta^T, \alpha)^T\}\Big].
\ese
Moving all terms without $(\bbeta, \alpha)$ to the left-hand side, we find:
\bse
&&\sum_{i=1}^n\Big\{\Delta_i (\bZ^a_i,\bs_i-X_i\bone_{m_i})^T(\bI_m - \bP_{\bZ^b_i})+\\
&&(1 - \Delta_i)(\bZ^a_i,\bs_i-\widehat{X}_i\bone_{m_i})^T(\bI_m - \bP_{(\bone_{m_i},\bZ^b_i)})\Big\}\bY_i\\
&=&\sum_{i=1}^n\Big\{\Delta_i (\bZ^a_i,\bs_i-X_i\bone_{m_i})^T(\bI_m - \bP_{\bZ^b_i})(\bZ^a_i,\bs_i-X_i\bone_{m_i})+\\
&&(1 - \Delta_i)(\bZ^a_i,\bs_i-\widehat{X}_i\bone_{m_i})^T(\bI_m - \bP_{(\bone_{m_i},\bZ^b_i)})(\bZ^a_i,\bs_i-\widehat{X}_i\bone_{m_i})\Big\}(\bbeta^T, \alpha)^T.
\ese
From this, we obtain the following regression coefficient estimates:
\bse
(\hat{\bbeta}^T,\hat{\alpha})^T&=&\Big[\sum_{i=1}^n\Big\{\Delta_i (\bZ^a_i,\bs_i-X_i\bone_{m_i})^T(\bI_m - \bP_{\bZ^b_i})(\bZ^a_i,\bs_i-X_i\bone_{m_i})+\\
&&(1 - \Delta_i)(\bZ^a_i,\bs_i-X_i\bone_{m_i})^T(\bI_m - \bP_{(\bone_{m_i},\bZ^b_i)})(\bZ^a_i,\bs_i-X_i\bone_{m_i})\Big\}\Big]^{-1}\times\\
&&\Big[\sum_{i=1}^n\Big\{\Delta_i (\bZ^a_i,\bs_i-X_i\bone_{m_i})^T(\bI_m - \bP_{\bZ^b_i})\bY_i+\\
&&(1 - \Delta_i)(\bZ^a_i,\bs_i-X_i\bone_{m_i})^T(\bI_m - \bP_{\bZ^b_i})\bY_i\Big\}\Big]\\
&=&\bM_i^{-1}\Big[\sum_{i=1}^n\Big\{\Delta_i (\bZ^a_i,\bs_i-X_i\bone_{m_i})^T(\bI_m - \bP_{\bZ^b_i})\bY_i+\\
&&(1 - \Delta_i)(\bZ^a_i,\bs_i-X_i\bone_{m_i})^T(\bI_m - \bP_{\bZ^b_i})\bY_i\Big\}\Big],
\ese
assuming that the matrix 
\bse
\bM_i&=&\sum_{i=1}^n\Big\{\Delta_i (\bZ^a_i,\bs_i-X_i\bone_{m_i})^T(\bI_m - \bP_{\bZ^b_i})(\bZ^a_i,\bs_i-X_i\bone_{m_i})+\\
&&(1 - \Delta_i)(\bZ^a_i,\bs_i-X_i\bone_{m_i})^T(\bI_m - \bP_{(\bone_{m_i},\bZ^b_i)})(\bZ^a_i,\bs_i-X_i\bone_{m_i})\Big\}
\ese
is invertible.
Next, we consider the $(p+2)$th element of the estimating equation (i.e., the element corresponding to $\sigma^2$):
\bse
0&=&\sum_{i=1}^n\Bigg(\Delta_i\left[\frac{1}{2}\{\bY_i^T\bY_i-\E(\bY_i^T\bY_i|\bT_i,\tilD_i)\}-\bzeta^T(\tilD_i;\btheta)\{\bY_i-\E(\bY_i|\bT_i,\tilD_i)\}\right]+\\
&&(1-\Delta_i)\left[\frac{1}{2}\{\bY_i^T\bY_i-\E(\bY_i^T\bY_i|\bT_i,\widehat{\tilD}_i)\}-\bzeta^T(\widehat{\tilD}_i;\btheta)\{\bY_i-\E(\bY_i|\bT_i,\widehat{\tilD}_i)\}\right]\Bigg)\\
&=&\sum_{i=1}^n\Bigg(\Delta_i\Big[\frac{1}{2}\{\bY_i^T\bY_i-\sigma^2(m_i-q)+||\rm{E}(\bY_i|\bT_i,\tilD_i)||^2_2\}\\
&&-\bzeta^T(\tilD_i;\btheta)\{\bY_i-\E(\bY_i|\bT_i,\tilD_i)\}\Big]+\\
&&(1-\Delta_i)\Big[\frac{1}{2}\{\bY_i^T\bY_i-\sigma^2(m_i-q)+||\rm{E}(\bY_i|\bT_i^{\rm aug},\tilD^*)||^2_2\}-\\
&&\bzeta^T(\widehat{\tilD}_i;\btheta)\{\bY_i-\E(\bY_i|\bT_i,\widehat{\tilD}_i)\}\Big]\Bigg).
\ese
From this, we see that the estimate of $\sigma^2$ is given by:
\bse
\hat{\sigma^2}&=&\left[\sum_{i=1}^n\left\{\Delta_i(m_i-q)+(1-\Delta_i)(m_i-q)\right\}\right]^{-1} \times\\
&&\Bigg\{\sum_{i=1}^n\Bigg(\Delta_i\left[\frac{1}{2}\{\bY_i^T\bY_i+||\rm{E}(\bY_i|\bT_i,\tilD_i)||^2_2\}-\bzeta^T(\tilD_i;\btheta)\{\bY_i-\E(\bY_i|\bT_i,\tilD_i)\}\right]+\\
&&(1-\Delta_i)\left[\frac{1}{2}\{\bY_i^T\bY_i+||\rm{E}(\bY_i|\bT_i^{\rm aug},\tilD^*)||^2_2\}-\bzeta^T(\widehat{\tilD}_i;\btheta)\{\bY_i-\E(\bY_i|\bT_i,\widehat{\tilD}_i)\}\right]\Bigg)\Bigg\}.
\ese
With this, we have established the identifiability of $\btheta$.
\section{Consistency}
\label{sec:consistency_proof}
We now turn our attention to the consistency of our novel estimator, $\ACEest$ (Theorem~\ref{theorem:consistency}). This theorem requires that we first prove that our full estimating function has mean $\bzero$ (Proposition~\ref{prop:mean_zero}). Yet we begin by proving Lemma~\ref{lem:cond_ind_claims}, which establishes an important relationship between assumptions (C1) and (C2) outlined in Proposition~\ref{prop:mean_zero} and the narrower assumptions required to prove that the full estimating function has mean $\bzero$.

\subsection{Relationship between conditional independence claims}
\label{sec:cond-ind-relationship}
\begin{Lem}
\label{lem:cond_ind_claims}
Consider the transformations $\bT_i=\sigma^2\bZ_i^{bT}\bY_i$ and $\bT_i^{\rm aug}=\sigma^{-2}(\bone_{m_i},\bZ^b_i)^T\bY_i$.
\begin{enumerate}
    \item[(A)] If $\bY_i$ is conditionally independent of $C_i$ given $\tilD_i$, then it must also be true that $\bY_i$ is conditionally independent of $C_i$ given $(\bT_i, \tilD_i)$.
    \item[(B)] If $\bY_i$ is conditionally independent of $C_i$ given $\widehat{\tilD}_i$, then it must also be true that $\bY_i$ is conditionally independent of $C_i$ given $(\bT_i^{\rm aug}, \widehat{\tilD}_i)$.
\end{enumerate}
\end{Lem}
We prove Lemma~\ref{lem:cond_ind_claims} (A) and omit the proof of Lemma~\ref{lem:cond_ind_claims} (B), since the proof of the latter is incredibly similar to that of the former (with $(\tilD_i, \bT_i)$ replaced by $(\widehat{\tilD}_i, \bT_i^{\rm aug})$). We first show that the premise in Lemma~\ref{lem:cond_ind_claims} (A) is equivalent to:
\bse
f_{\bY_i| \tilD_i, C_i}(\bY_i|\tilD_i,C_i) = f_{\bY_i| \tilD_i}(\bY_i|\tilD_i).
\ese
From this and Bayes theorem, we observe the following:
\bse
\frac{f_{\bY_i, \tilD_i| C_i}(\bY_i,\tilD_i|C_i)}{f_{\tilD_i| C_i}(\tilD_i|C_i)} = \frac{f_{\bY_i, \tilD_i}(\bY_i,\tilD_i)}{f_{ \tilD_i}(\tilD_i)}.
\ese
Then, since $\bT_i$ is a function of $\bY_i$ and $\tilD_i$, it follows that
\be
\label{eqn:cond-ind-with-W}
\frac{f_{\bY_i, \tilD_i, \bT_i| C_i}(\bY_i,\tilD_i, \bT_i|C_i)}{f_{\tilD_i| C_i}(\tilD_i|C_i)} = \frac{f_{\bY_i, \tilD_i, \bT_i}(\bY_i,\tilD_i,\bT_i)}{f_{ \tilD_i}(\tilD_i)}.
\ee
We now apply Bayes theorem to modify the denominator on the left hand side of Equation~\eqref{eqn:cond-ind-with-W}:
\bse
&\frac{f_{\tilD_i,\bT_i|C_i}(\tilD_i,\bT_i|C_i)}{f_{ \tilD_i|C_i}(\tilD_i|C_i)}=f_{\bT_i|\tilD_i,C_i}(\bT_i|\tilD_i,C_i) \\
\Rightarrow& f_{ \tilD_i|C_i}(\tilD_i|C_i)=\frac{f_{\tilD_i,\bT_i|C_i}(\tilD_i,\bT_i|C_i)}{f_{\bT_i|\tilD_i,C_i}(\bT_i|\tilD_i,C_i)}.
\ese
Then, because $\bY_i$ is conditionally independent of $C_i$ given $\tilD_i$ per our assumption in Lemma~\ref{lem:cond_ind_claims} (A), we know that $\bT_i$ is also conditionally independent of $C_i$ given $\tilD_i$ since $\bT_i$ is a function of $\bY_i$ and $\tilD_i$. Therefore, $f_{\bT_i|\tilD_i,C_i}(\bT_i|\tilD_i,C_i)=f_{\bT_i|\tilD_i}(\bT_i|\tilD_i)$ and
\bse
f_{ \tilD_i|C_i}(\tilD_i|C_i)=\frac{f_{\tilD_i,\bT_i|C_i}(\tilD_i,\bT_i|C_i)}{f_{\bT_i|\tilD_i}(\bT_i|\tilD_i)}.
\ese
We now apply Bayes theorem to modify the denominator on the right hand side of Equation~\eqref{eqn:cond-ind-with-W}:
\bse
&\frac{f_{\tilD_i,\bT_i}(\tilD_i,\bT_i)}{f_{ \tilD_i}(\tilD_i)}=f_{\bT_i|\tilD_i}(\bT_i|\tilD_i)\nonumber\\
\Rightarrow&f_{ \tilD_i}(\tilD_i)=\frac{f_{\tilD_i,\bT_i}(\tilD_i,\bT_i)}{f_{\bT_i|\tilD_i}(\bT_i|\tilD_i)}.
\ese
With these two modifications to the denominators of Equation~\eqref{eqn:cond-ind-with-W}, we obtain:
\bse
\frac{f_{\bY_i, \tilD_i, \bT_i| C_i}(\bY_i,\tilD_i, \bT_i|C_i)f_{\bT_i|\tilD_i}(\bT_i|\tilD_i)}{f_{\tilD_i,\bT_i|C_i}(\tilD_i,\bT_i|C_i)} = \frac{f_{\bY_i, \tilD_i, \bT_i}(\bY_i,\tilD_i,\bT_i)f_{\bT_i|\tilD_i}(\bT_i|\tilD_i)}{f_{\tilD_i,\bT_i}(\tilD_i,\bT_i)}.
\ese
Then, we see that the conditional density $f_{\bT_i|\tilD_i}(\bT_i|\tilD_i)$ cancels from both sides to yield:
\bse
\frac{f_{\bY_i, \tilD_i, \bT_i| C_i}(\bY_i,\tilD_i, \bT_i|C_i)}{f_{\tilD_i,\bT_i|C_i}(\tilD_i,\bT_i|C_i)} = \frac{f_{\bY_i, \tilD_i, \bT_i}(\bY_i,\tilD_i,\bT_i)}{f_{\tilD_i,\bT_i}(\tilD_i,\bT_i)},
\ese
which is equivalent to the statement that $f_{\bY_i|\tilD_i,\bT_i,C_i}(\bY_i|\tilD_i,\bT_i,C_i)=f_{\bY_i|\tilD_i,\bT_i}(\bY_i,\tilD_i,\bT_i)$ and hence that $\bY_i$ is conditionally independent of $C_i$ given $(\tilD_i, \bT_i)$. This confirms Lemma~\ref{lem:cond_ind_claims} (A); the proof of Lemma~\ref{lem:cond_ind_claims} (B) proceeds analogously and so is omitted for brevity.

\subsection{Proof of Proposition~\ref{prop:mean_zero}}
\label{sec:mean_zero_proof}
% \begin{proposition}
% \label{prop:mean_zero}
% Consider the efficient score vectors $\bS_{\rm{eff}}(\bY_i,\tilD_i;\btheta)$ and $\bS_{\rm{eff}}^*(\bY_i,\widehat{\tilD}_i;\btheta)$ defined in Equations~\eqref{eqn:app_uncens_eff_score_vec} and \eqref{eqn:app_cens_eff_score_vec}, respectively, and let $\Delta_i = I(X_i \leq C_i)$ be the censoring indicator. 
% \begin{enumerate}
%     \item[(A)] If $\bY_i$ is conditionally independent of $C_i$ given $(\tilD_i, \bT_i)$, then it is also true that $\E\{\Delta_i\bS_{\rm{eff}}(\bY_i,\tilD_i;\btheta)\} = \bzero$.
%     \item[(B)] If $\bY_i$ is conditionally independent of $C_i$ given $(\widehat{\tilD}_i, \bT_i^{\rm aug})$, then it is also true that $\E\{(1-\Delta_i)\bS_{\rm{eff}}^*(\bY_i,\widehat{\tilD}_i;\btheta)\} = \bzero$.
% \end{enumerate}
% \end{proposition}

We prove Proposition~\ref{prop:mean_zero} (A) and omit the proof of Proposition~\ref{prop:mean_zero} (B), since the proof of the latter follows analogously from that of the former (with $(\tilD_i, \bT_i)$ replaced by $(\widehat{\tilD}_i, \bT_i^{\rm aug})$). 
% Moreover, per Lemma~\ref{lem:cond_ind_claims}, the premise of Proposition~\ref{prop:mean_zero} (A) implies that $\bY_i$ is conditionally independent of $C_i$ given $(\tilD_i, \bT_i)$.
To prove Proposition~\ref{prop:mean_zero} (A), we begin by applying the law of total expectation to find:
\bse
\E\{\Delta_i\bS_{\rm{eff}}(\bY_i,\tilD_i;\btheta)\}&=&\E[\E\{\Delta_i\bS_{\rm{eff}}(\bY_i,\tilD_i;\btheta)|\bT_i,\tilD_i,C_i\}]\\
&=&\E\{I(X_i \leq C_i)\bS_{\rm{eff}}(\bY_i,\tilD_i;\btheta)|\bT_i,\tilD_i,C_i\}
\\
&=&I(X_i \leq C_i)\E\{\bS_{\rm{eff}}(\bY_i,\tilD_i;\btheta)|\bT_i,\tilD_i,C_i\}
\ese
since, by definition, $\tilD_i=(\bZ^a_i, \bs_i, X_i, \bZ^b_i)$ includes $X_i$. Hence, to show the desired result that $\E\{\Delta_i\bS_{\rm{eff}}(\bY_i,\tilD_i;\btheta)\} = \bzero$, it suffices to show that $\E\{\bS_{\rm{eff}}(\bY_i,\tilD_i;\btheta)|\bT_i,\tilD_i,C_i\}=\bzero$. We accomplish this in a component-wise fashion. Recall from Section~\ref{sec:beta_uncens} that:
\bse
\bS_{\rm{eff},\bbeta}(\bY_i,\tilD_i;\btheta)
=\frac{\bZ^{aT}_i\{\bY_i-\E(\bY_i|\bT_i,\tilD_i)\}}{\sigma^2}.
\ese
Therefore,
\bse
\E\{\bS_{\rm{eff,\bbeta}}(\bY_i,\tilD_i)|\bT_i,\tilD_i,C_i\}&=&\frac{\E[\bZ^{aT}_i\{\bY_i-\E(\bY_i|\bT_i,\tilD_i)\}|\bT_i,\tilD_i,C_i]}{\sigma^2}\\
&=&\frac{\bZ^{aT}_i\{\E(\bY_i|\bT_i,\tilD_i,C_i)-\E(\bY_i|\bT_i,\tilD_i)\}}{\sigma^2}.
\ese
Per Lemma~\ref{lem:cond_ind_claims}, our assumption that $\bY_i$ is conditionally independent of $C_i$ given $\tilD_i$ implies the conditional independence of $\bY_i$ and $C_i$ given $(\bT_i,\tilD_i)$. It therefore follows that $E(\bY_i|\bT_i,\tilD_i,C_i)=E(\bY_i|\bT_i,\tilD_i)$ and hence that $E\{\bS_{\rm{eff},\bbeta}(\bY_i,\tilD_i;\btheta)|\bT_i,\tilD_i,C_i\}=\bzero$.

The proof corresponding to $S_{\rm{eff},\alpha}(\bY_i,\tilD_i;\btheta)$ proceeds similarly. From Section~\ref{sec:alpha_uncens}, we know that for uncensored subjects, the efficient score vector corresponding to $\alpha$ is:
\bse
S_{\rm{eff},\alpha}(\bY_i,\tilD_i;\btheta)&=&\frac{(\bs_i-X_i\bone_{m_i})^T\{\bY_i-\E(\bY_i|\bT_i,\tilD_i)\}}{\sigma^2}.
\ese
As a result, we observe that:
\bse
\E\{\bS_{\rm{eff,\alpha}}(\bY_i,\tilD_i)|\bT_i,\tilD_i,C_i\}&=&\frac{(\bs_i-X_i\bone_{m_i})^T\{\E(\bY_i|\bT_i,\tilD_i,C_i)-\E(\bY_i|\bT_i,\tilD_i)\}}{\sigma^2}.
\ese
Again, we know from our assumption --- that $\bY_i$ is conditionally independent of $C_i$ given $\tilD_i$ --- and Lemma~\ref{lem:cond_ind_claims} that $\E\{\bS_{\rm{eff,\alpha}}(\bY_i,\tilD_i)|\bT_i,\tilD_i,C_i\}=0$.

Next, recall from Section~\ref{sec:sigma_uncens} that: 
\bse
S_{\rm{eff},\sigma^2}(\bY_i,\tilD_i;\btheta)=\frac{\frac{1}{2}\left\{\bY_i^T\bY_i-\E(\bY_i^T\bY_i|\bT_i,\tilD_i)\right\}-\bzeta(\tilD_i;\btheta)^T\{\bY_i-\E(\bY_i|\bT_i,\tilD_i)\}}{\sigma^4}.
\ese
From this, we know that
\bse
\sigma^4\E\{\bS_{\rm{eff,\sigma^2}}(\bY_i,\tilD_i)|\bT_i,\tilD_i,C_i\}&=&\frac{1}{2}\left\{\E(\bY_i^T\bY_i|\bT_i,\tilD_i,C_i)-\E(\bY_i^T\bY_i|\bT_i,\tilD_i)\right\}-\\
&&\bzeta(\tilD_i;\btheta)^T\{\E(\bY_i|\bT_i,\tilD_i,C_i)-\E(\bY_i|\bT_i,\tilD_i)\}
\ese
which, similarly, is $0$ because we assume $\bY_i$ is conditionally independent of $C_i$ given $\tilD_i$. This establishes that each component of $\bS_{\rm eff}(\bY_i,\tilD_i)$ has mean $\bzero$, confirming Proposition~\ref{prop:mean_zero} (A). As noted, the proof of  Proposition~\ref{prop:mean_zero} (B) proceeds in parallel (with $(\tilD_i, \bT_i)$ replaced by $(\widehat{\tilD}_i, \bT_i^{\rm aug})$).

\section{Proof of Theorem~\ref{theorem:asymp_norm}}
\label{sec:asymp_norm_proof}

Again, we use $\bS_{\rm{eff}}^{\rm{full}}(Y_i,\tilD_i,\widehat{\tilD}_i;\btheta) = \Delta_i\bS_{\rm{eff}}(\bY_i,\tilD_i;\btheta) + (1 - \Delta_i)\bS_{\rm{eff}}^*(\bY_i,\widehat{\tilD}_i;\btheta)$ to denote the full estimating function. Since $\E\left\{\bS_{\rm{eff}}^{\rm{full}}(Y_i,\tilD_i,\widehat{\tilD}_i;\btheta)\right\}=\bzero$ under conditions (C1) and (C2) in Proposition~\ref{prop:mean_zero}, we can use Taylor's theorem to expand around $\btheta_0$ as follows:
\bse
\bzero&=& \sum_{i=1}^n \left\{\bS_{\rm{eff}}^{\rm{full}}(Y_i,\tilD_i,\widehat{\tilD}_i;\ACEest)\right\}\\
&=& n^{-1/2}\sum_{i=1}^n \left\{\bS_{\rm{eff}}^{\rm{full}}(Y_i,\tilD_i,\widehat{\tilD}_i;\btheta_0)\right\} +\\
&&n^{-1}\sum_{i=1}^n \left\{\frac{\partial}{\partial \btheta^T} \bS_{\rm{eff}}^{\rm{full}}(Y_i,\tilD_i,\widehat{\tilD}_i;\btheta^*)\right\}n^{1/2}(\ACEest-\btheta_0),
\ese
where $\btheta^*$ lies on the line connecting $\ACEest$ and $\btheta_0$. Then, by C6,
\bse
\bzero&=& n^{-1/2}\sum_{i=1}^n \left\{\bS_{\rm{eff}}^{\rm{full}}(Y_i,\tilD_i,\widehat{\tilD}_i;\btheta_0)\right\} + \\
&&\E\left\{\frac{\partial}{\partial \btheta^T} \bS_{\rm{eff}}^{\rm{full}}(Y_i,\tilD_i,\widehat{\tilD}_i;\btheta_0)\right\}n^{1/2}(\ACEest-\btheta_0) + o_p(1).
\ese
Therefore, 
\bse
n^{1/2}(\ACEest-\btheta_0)&=&\left[\E\left\{\frac{\partial}{\partial \btheta^T} \bS_{\rm{eff}}^{\rm{full}}(Y_i,\tilD_i,\widehat{\tilD}_i;\btheta_0)\right\}\right]^{-1}\times\\
&&n^{-1/2}\sum_{i=1}^n \bS_{\rm{eff}}^{\rm{full}}(Y_i,\tilD_i,\widehat{\tilD}_i;\btheta_0)  + o_p(1)
\ese
by C7. Hence, it follows directly from the central limit theorem that $n^{1/2}(\ACEest-\btheta_0)$ converges in distribution to $\Normal_{p_a+2}(0, {\cal V})$, where ${\cal V} = {\cal A}^{-1} {\cal B} ({\cal A}^{-1})^T$ with 
\bse
{\cal A}^{-1}&=& \left[\E\left\{\frac{\partial}{\partial \btheta^T} \bS_{\rm{eff}}^{\rm{full}}(Y_i,\tilD_i,\widehat{\tilD}_i;\btheta_0)\right\}\right]^{-1}\\
{\cal B}&=&  \Var\{\bS_{\rm{eff}}^{\rm{full}}(Y_i,\tilD_i,\widehat{\tilD}_i;\btheta_0)\}.
\ese
\section{Limitation}
\label{sec:limitation}
We now demonstrate a limitation with the proposed estimating equation:
\begin{Lem}
\label{lem:limitation}
Let $C(\bM)$ denote the column space of a matrix $\bM$ and let $(\bZ^a_i)_k$ denote the $k$th column of $\bZ^a_i$. 
\begin{enumerate}
    \item[A] If $(\bZ^a_i)_k$ belongs to $C(\bZ^b_i)$, then the $k$th element of $S_{\rm{eff},\bbeta}(\bY_i,\tilD_i;\btheta)$ will be $0$. Similarly, if $(\bs_i-X_i\bone_{m_i})$ belongs to $C(\bZ^b_i)$, then $S_{\rm{eff},\alpha}(\bY_i,\tilD_i;\btheta)$ will be $0$.
    \item[B] If $(\bZ^a_i)_k$ belongs to $C\{(\bone_{m_i},\bZ_i^b)\}$, then the $k$th element of $S_{\rm{eff},\bbeta}^*(\bY_i,\widehat{\tilD}_i;\btheta)$ will be $0$. Similarly, if $(\bs_i-\widehat{X}_i\bone_{m_i})$ belongs to $C\{(\bone_{m_i},\bZ_i^b)\}$, then $S_{\rm{eff},\alpha}^*(\bY_i,\widehat{\tilD}_i;\btheta)$ will be $0$.
\end{enumerate}
\end{Lem}

Lemma~\ref{lem:limitation} follows quickly from the definitions of the efficient score vectors given in Section~\ref{sec:calcs}. To prove this, first suppose that $(\bZ^a_i)_k$ belongs to the column space of $\bZ_i^b$. Then, because $\bP_{\bZ_i^b}$ is the orthogonal projection operator onto the column space of $\bZ_i^b$, we know that $\bI_{m_i}(\bZ^a_i)_k=(\bZ^a_i)_k =\bP_{\bZ^b_i}(\bZ^a_i)_k$. This implies that $(\bI_{m_i} - \bP_{\bZ^b_i})(\bZ^a_i)_k = \bzero$ and, equivalently, that $(\bZ^a_i)_k^T(\bI_{m_i} - \bP_{\bZ^b_i}) = \bzero^T$. As a result, $\bZ^{aT}_i(\bI_{m_i} - \bP_{\bZ^b_i})$ has $k$th row equal to $\bzero$ and hence, the $k$th element of $S_{\rm{eff},\bbeta}(\bY_i,\tilD_i;\btheta)$ will be 0. Proposition~\ref{prop:limitation} --- an immediate consequence of Lemma~\ref{lem:limitation} --- describes the broader situation in which this poses a computational challenge.

\begin{proposition}
\label{prop:limitation}
Let $k$ be a fixed integer between $1$ and $p_a$. Suppose that $(\bZ^a_i)_k$ belongs to $C(\bZ^b_i)$ for all $i$ such that $\Delta_i=1$. Further suppose that $(\bZ^a_i)_k$ belongs to $C\{(\bone_{m_i},\bZ_i^b)\}$ for all $i$ such that $\Delta_i=0$. Then, the $k$th element of the total estimating equation given in Equation~\eqref{eqn:total_estimating_equation} will be $0$.
\end{proposition}

In less generality, if $\bZ^a_i$ and $\bZ^b_i$ ``share'' any columns for all $i$, the estimating equation may fail to converge. This can happen if, for example, we include both a fixed and a random intercept in the outcome model of interest.

\section{Supplemental Tables}
\label{sec:app_tables}

% latex table generated in R 4.1.0 by xtable 1.8-4 package
% Thu Nov 17 11:12:55 2022
\begin{table}[hbt!]
\centering
\begin{tabular}{lllrrrrr}
  \hline
Censoring & Parameter & Method & Bias & SEE & ESE & MSE & CPr \\ 
  \hline
Light & $\alpha$ & ACE & -0.000 & 0.022 & 0.022 & 0.000 & 0.950 \\ 
  % &  & CMI & 0.197 & 0.022 & 0.132 & 0.056 & 0.046 \\ 
  &  & MCMI & 0.173 & 0.023 & 0.127 & 0.046 & 0.070 \\ 
  & $\beta$ & ACE & 0.000 & 0.024 & 0.024 & 0.001 & 0.942 \\ 
  % &  & CMI & 0.002 & 0.036 & 0.037 & 0.001 & 0.957 \\ 
  &  & MCMI & 0.002 & 0.036 & 0.038 & 0.001 & 0.951 \\ 
  & $\sigma^2$ & ACE & -0.003 & 0.034 & 0.034 & 0.001 & 0.940 \\ 
  % &  & CMI & 1.819 &  & 2.911 & 11.771 &  \\ 
  &  & MCMI & 1.940 &  & 2.979 & 12.629 &  \\
   \hline
\end{tabular}
\caption{Simulation results under light censoring when the imputation model is correctly specified. ACE: proposed estimating equation applied to the imputed dataset. MCMI: multiple conditional mean imputation. We present empirical bias, the average standard error estimates (SEE), the empirical standard error (ESE), the average mean squared errors (MSE), and the observed coverage probability (CPr) of the Wald-type confidence intervals with nominal 95\% coverage.}
\label{table:corr_spec_results_light}
\end{table}

\begin{table}[hbt!]
\footnotesize
\centering
\begin{tabular}{lllrrrrr}
  \hline
Censoring & Parameter & Method & Bias & SEE & ESE & MSE & Coverage \\ 
  \hline
  Light & $\alpha$ & ACE & 0.000 & 0.022 & 0.022 & 0.000 & 0.946 \\ 
  % &  & CMI & 1.874 & 0.173 & 10.452 & 112.645 & 0.000 \\ 
  &  & MCMI & 1.872 & 0.175 & 10.429 & 112.154 & 0.000 \\ 
  & $\beta$ & ACE & 0.001 & 0.024 & 0.024 & 0.001 & 0.948 \\ 
  % &  & CMI & -0.071 & 0.316 & 1.287 & 1.660 & 0.953 \\ 
  &  & MCMI & -0.071 & 0.316 & 1.289 & 1.665 & 0.953 \\ 
  & $\sigma^2$ & ACE & -0.002 & 0.034 & 0.035 & 0.001 & 0.944 \\ 
  % &  & CMI & $\geq 100$ &  & $\geq 100$ & $\geq 100$ &  \\ 
  &  & MCMI & $\geq 100$ &  & $\geq 100$ & $\geq 100$ &  \\
   \hline
\end{tabular}
\caption{Simulation results under light censoring when the imputation model is mis-specified. ACE: proposed estimating equation applied to the imputed dataset. MCMI: multiple conditional mean imputation. We present empirical bias, the average standard error estimates (SEE), the empirical standard error (ESE), the average mean squared errors (MSE), and the observed coverage probability (CPr) of the Wald-type confidence intervals with nominal 95\% coverage.}
\label{table:mis_spec_results_light}
\end{table}

% latex table generated in R 4.2.0 by xtable 1.8-4 package
% Tue Aug 16 17:29:53 2022
\begin{table}[ht]
\footnotesize
\centering
\begin{tabular}{lll}
  \hline
 & Censored ($\Delta_i$ = 0) & Uncensored ($\Delta_i$ = 1)\\ 
  \hline
n & 858 & 244 \\ 
  Age & 38.82 (10.43) & 43.23 (10.26) \\ 
  Sex = Male (\%) & 311 (36.2) & 83 (34.0) \\ 
  Education & 14.64 (2.64) & 14.11 (2.55) \\ 
  CAG & 42.22 (2.62) & 43.48 (2.81) \\ 
  CAG-Age Product & 304.68 (78.74) & 389.54 (72.88) \\ 
  cUHDRS & 16.88 (1.78) & 15.35 (1.94) \\ 
  Total Motor Score & 3.79 (4.27) & 8.36 (6.71) \\ 
  SDMT Score & 52.47 (11.14) & 44.57 (10.56) \\ 
  Total Functional Capacity & 12.83 (0.75) & 12.70 (0.80) \\ 
  Stroop Color Test Score & 79.20 (13.69) & 70.95 (13.40) \\ 
  Stroop Word Test Score & 101.01 (17.28) & 91.62 (16.38) \\ 
  Stroop Interference Test Score & 46.45 (10.25) & 39.55 (9.16) \\ 
  $\widehat{X}_i$ & 11.26 (1.35) & 4.50 (2.78) \\
   \hline
\end{tabular}
\caption{Summary of model covariates ($\bZ^a_{ij}$) and potential outcomes ($\bY_i$) at visit 1. CAG: number of cytosine-adenine-guanine repeats. cUHDRS: composite Unified Huntington Disease Rating Scale. SDMT: Symbol Digits Modality Test.  $\widehat{X}_i=\Delta_iX_i+(1-\Delta_i){\rm E}(X_i|X_i>C_i,\bV_i)$. Continuous variables summarized by mean (standard deviation). Binary variables summarized by count (percent).}
\label{table:TableOne}
\end{table}

\begin{figure}
    \centering
    \includegraphics[width=0.7\textwidth]{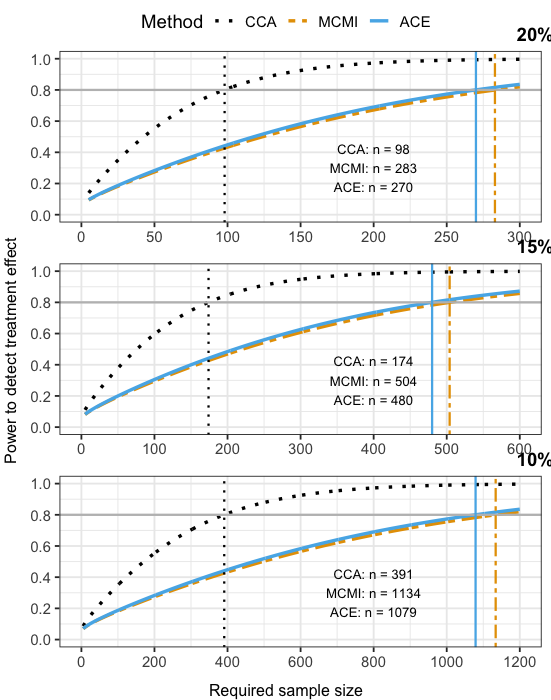}
    \caption{Power to detect 10\%, 15\%, and 20\% treatment effects for given per-group sample sizes with type 1 error rate $0.05$. Results are based on estimated slope in placebo group from complete case analysis (CCA), multiple conditional mean imputation (MCMI), and ACE imputation (ACE).}
    \label{fig:power_curves}
    % take out the T* plots, remove grid, are the numbers readable?, move the T* plots to later section, think about coloring, remove color (needs to be black and white ready)
\end{figure}

\end{document}